\documentstyle[preprint,aps,eqsecnum,epsfig]{revtex}
\tighten
\begin{document}
\preprint{PITT-97-677; CMU-HEP-97-08; LPTHE-97; DOE-ER/40682-133} 
\draft
\title{\bf NON-PERTURBATIVE QUANTUM DYNAMICS OF A NEW INFLATION MODEL}  
\author{{\bf D. Boyanovsky$^{(a)}$, D. Cormier$^{(b)}$,
H. J. de Vega$^{(c)}$, R. Holman$^{(b)}$, S. P. Kumar$^{(b)}$}}
\address
{(a)  Department of Physics and Astronomy, University of
Pittsburgh, Pittsburgh, PA. 15260, U.S.A. \\
 (b) Department of Physics, Carnegie Mellon University, Pittsburgh,
PA. 15213, U. S. A. \\
 (c)  LPTHE, \footnote{Laboratoire Associ\'{e} au CNRS UA280.}
Universit\'e Pierre et Marie Curie (Paris VI) 
et Denis Diderot  (Paris VII), Tour 16, 1er. \'etage, 4, Place Jussieu
75252 Paris, Cedex 05, France}
\date{September 1997}
\maketitle
\begin{abstract}
 We consider an $ O(N) $ model coupled self-consistently to gravity in the
semiclassical approximation, where the field is subject to `new inflation'
type initial conditions. We study the dynamics  self-consistently and
non-perturbatively with non-equilibrium field theory methods in
the large $ N $ limit. We find that spinodal instabilities drive the growth
of non-perturbatively large quantum fluctuations which shut off the
inflationary growth of the scale factor. We find that 
a very specific combination of  these large
fluctuations plus the inflaton zero mode assemble into a new effective field.
This new field behaves classically and it is the object which actually
rolls down.
We show how  this 
reinterpretation saves the standard picture of how metric perturbations are
generated during inflation and that the spinodal growth of fluctuations
dominates the time dependence of the Bardeen variable for superhorizon modes
during inflation.  We compute the amplitude and index for the spectrum of
scalar density and tensor perturbations and argue that in all models of this
type the spinodal instabilities are responsible for a `red' spectrum of
primordial scalar density perturbations.  A criterion for the validity of these
models is provided and contact with the reconstruction program is established
validating some of the results within a non-perturbative framework.  The
decoherence aspects and the quantum to classical transition through inflation
are studied in detail by following the full evolution of the density matrix and
relating the classicality of cosmological perturbations to that of
long-wavelength matter fluctuations.
 
\end{abstract}

\section{Introduction and Motivation}

Inflationary cosmology has come of age. From its beginnings as a solution to
purely theoretical problems such as the horizon, flatness and monopole
problems\cite{guth}, it has grown into the main contender for the source of
primordial fluctuations giving rise to large scale
structure\cite{turok,albrecht,dodelsondef}. There is evidence from the
measurements of temperature anisotropies in the cosmic microwave background
radiation (CMBR) that the scale invariant power spectrum predicted by generic
inflationary models is at least consistent with
observations\cite{kolb,turner,liddle,lyth1,smoot1} and we can expect further
and more exacting tests of the inflationary power spectrum when the MAP and
PLANCK missions are flown. In particular, if the fluctuations that are
responsible for the temperature anisotropies of the CMB truly originate from
quantum fluctuations during inflation, determinations of the spectrum of scalar
and tensor perturbations will constrain inflationary models based on particle
physics scenarios and probably will validate or rule out specific
proposals\cite{turner,liddle,lyth2,dodelson}. Already current bounds on the
spectrum of scalar density perturbations seem to rule out some versions of
`extended' inflation\cite{lyth2}.

The tasks for inflationary universe enthusiasts are then two-fold. First,
models of inflation must be constructed that have a well-defined rationale in
terms of coming from a reasonable particle physics model. This is in contrast
to the current situation where most, if not all acceptable inflationary models
are ad-hoc in nature, with fields and potentials put in for the sole purpose of
generating an inflationary epoch. Second, and equally important, we must be
sure that the quantum dynamics of inflation is well understood. This is
extremely important, especially in light of the fact that it is {\it exactly} this
quantum behavior that is supposed to give rise to the primordial metric
perturbations which presumably have imprinted themselves in the CMBR. This
latter problem is the focus of this paper. 

The inflaton must be treated as a {\it non-equilibrium} quantum field . The
simplest way to see this comes from the requirement of having small enough
metric perturbation amplitudes which in turn requires that the quartic self
coupling 
$ \lambda $ of the inflaton be extremely small, typically of order $ \sim
10^{-12} $. Such a small coupling cannot establish local thermodynamic
equilibrium (LTE) for {\it all} field modes; near a phase transition the long wavelength modes will respond too slowly to be able to enter LTE. In fact, the
superhorizon sized modes will be out of the region of causal contact and
cannot thermalize. We see then that if we
want to gain a deeper understanding of inflation, non-equilibrium tools must be
developed. Such tools exist and have now been developed to the point that they
can give quantitative answers to these questions in 
cosmology\cite{us1,frwpaper,De Sitter,boylee,frw2}. These methods permit
us to follow the {\bf dynamics} of quantum fields in situations where the energy 
density is non-perturbatively large ($ \sim 1/\lambda $). That is, they allow
the computation of the time evolution of non-stationary states and of non-thermal 
density matrices.

Our approach is to apply non-equilibrium quantum field theory techniques to the
situation of a scalar field coupled to {\it semiclassical} gravity, where the
source of the gravitational field is the expectation value of the stress energy
tensor in the relevant, dynamically changing, quantum state. In this way
we can go beyond the standard analyses\cite{linde2,vilenkin,stein,guthpi} which
treat the background as fixed. 

We will mainly deal with `new inflation' scenarios, where a scalar field
$\phi$ evolves under the action of a typical symmetry breaking potential. The
initial conditions will be taken so that the initial value of the order 
parameter (the field expectation value) is near the top of the potential
(the disordered state) with essentially zero time derivative. 

What we find is that the existence of spinodal instabilities, i.e. the fact
that eventually (in an expanding universe) all modes will act as if they have a
{\it negative} mass squared, drives the quantum fluctuations to grow {\it
non-perturbatively} large. We have the picture of an initial wave-function or
density matrix peaked near the unstable state and then spreading until it
samples the stable vacua. Since these vacua are non-perturbatively far from
the initial state (typically $\sim m\slash \sqrt{\lambda}$, where $m$ is the
mass scale of the field and $\lambda$ the quartic self-coupling), the spinodal
instabilities will persist until the quantum fluctuations, as encoded in the
equal time two-point function $\langle \Phi(\vec{x}, t)^2 \rangle$, grow to
${\cal O}( m^2\slash \lambda$).

This growth eventually shuts off the inflationary behavior of the scale factor
as well as the growth of the quantum fluctuations (this last also happens
 in Minkowski spacetime \cite{us1}).

The scenario envisaged here is that of a quenched or supercooled phase 
transition where the order parameter is zero or very small. Therefore one is 
led to ask: 

a) What is rolling down?. 

b) Since the quantum fluctuations are non-perturbatively large ( $ \sim  
1/\lambda $), will not they modify drastically the FRW dynamics?.

c) How can one extract (small?) metric perturbations from non-perturbatively 
large field fluctuations?


We address the questions a)-c) as well as  other issues below. 

\section{Non-Equilibrium Quantum Field Theory, Semiclassical Gravity and
Inflation} 

Our program consists of finding ways to incorporate the non-equilibrium
behavior of the quantum fields involved in inflation into a framework that
treats gravity self consistently, at least in some approximation. We do this
via the use of semiclassical gravity\cite{birriel} where we say that
the metric is classical, at least to first approximation, whose source is
the expectation value of the stress energy tensor $\langle T_{\mu \nu} \rangle$
where this expectation value is taken in the dynamically determined state
described by the density matrix $\rho(t)$. This dynamical problem can be
described schematically as follows: 

\begin{enumerate}

\item{The dynamics of the scale factor $a(t)$ is driven by the 
semiclassical Einstein equations 

\begin{equation}
\frac{1}{8\pi G_R} G_{\mu \nu} + \frac{\Lambda_R}{8\pi G_R} g_{\mu \nu} +
\left(\rm{higher\  curvature}\right)= -\langle T_{\mu \nu}\rangle_R .
\end{equation}
Here $G_R, \Lambda_R$ are the renormalized values of Newton's constant and the
cosmological constant, respectively and $G_{\mu \nu}$ is the Einstein
tensor. The higher curvature terms must be included to absorb divergences.}

\item{On the other hand, the density matrix $\rho(t)$ that determines $\langle
T_{\mu \nu}\rangle_R$ obeys the Liouville equation

\begin{equation}
i\frac{\partial \rho(t)}{\partial t} = \left[H, \rho(t)\right],
\end{equation}
where $H$ is the evolution Hamiltonian, which is dependent on the scale factor,
 $a(t)$.}
\end{enumerate}

It is this set of equations we must try to solve; it is clear that initial
conditions must be appended to these equations for us to be able to arrive at
unique solutions to them. Let us discuss some aspects of the initial state of
the field theory first.

\subsection{On the initial state: dynamics of phase transitions} 

As we mentioned above, the situation we consider is one in which the theory
admits a symmetry breaking potential and in which the field expectation value
starts its evolution near the unstable point. There is an
issue as to how the field got to have an expectation value
near the unstable point
(typically at $\Phi=0$) as well as an issue concerning the initial state of the
non-zero momentum modes. The issue of initial conditions is present in any
formulation of inflation but chaotic. 

Since our background is an FRW spacetime, it is spatially homogeneous and we
can choose our state $\rho(t)$ to respect this symmetry. Starting from the full
quantum field $\Phi(\vec{x},t)$ we can extract a part that has a natural
interpretation as the zero momentum, c-number part of the field by writing:

\begin{eqnarray}
\Phi(\vec{x}, t)& = &\phi(t) + \Psi(\vec{x}, t) \nonumber \\ \phi(t) & = & {\rm
Tr}[\rho(t)\Phi(\vec{x}, t)]\equiv \langle \Phi(\vec{x}, t)) \rangle.
\end{eqnarray}
The quantity $\Psi(\vec{x}, t)$ represents the quantum fluctuations about the
zero mode $\phi(t)$ and clearly satisfies $\langle \Psi(\vec{x}, t) \rangle=0$.

We need to choose a basis to represent the density matrix. A natural choice
consistent with the translational invariance of our quantum state is that given
by the Fourier modes, in {\it comoving} momentum space, of the quantum
fluctuations $\Psi(\vec{x}, t)$:

\begin{equation}
\Psi(\vec{x}, t) = \int \frac{d^3 k}{(2 \pi)^3} \exp(-i\ \vec{k} \cdot \vec{x})\
\psi_k (t) .
\end{equation}

In this language we can state our ansatz for the initial condition of the
quantum state as follows. We take the zero mode $\phi(t=0)=\phi_0,\
\dot{\phi}(t=0)=0$, where $\phi_0$ will typically be very near the origin,
while the initial conditions on the the nonzero modes $\psi_k (t=0)$ will be
chosen such that the initial density matrix $\rho(t=0)$ describes a vacuum state
(i.e. an initial state in local thermal equilibrium at a temperature $T_i=0$).
There are some subtleties involved in this choice. First, as
explained in \cite{frw2}, in order for the density matrix to commute with the
initial Hamiltonian, we must choose the modes to be initially in the conformal
adiabatic vacuum (these statements will be made more precise below). This
choice has the added benefit of allowing for time independent renormalization
counterterms to be used in renormalizing the theory.

We are making the assumption of an initial vacuum state
in order to be able to proceed with the calculation. It would be
interesting to understand what forms of the density matrix can be used for
other, perhaps more reasonable, initial conditions. 

The assumptions of an initial equilibrium vacuum state are essentially 
the same used by Linde\cite{linde2},
Vilenkin\cite{vilenkin} , as well as by Guth and Pi\cite{guthpi} in their
analyses of the quantum mechanics of inflation in a fixed de Sitter
background. 

As discussed in the introduction, if we start from such an initial state,
spinodal instabilities will drive the growth of non-perturbatively large
quantum fluctuations. In order to deal with these, we need to be able to
perform calculations that take these large fluctuations into account. Although
the quantitative features of the dynamics will depend on the initial state,
the qualitative features associated with spinodal instabilities will be fairly 
robust for a wide choice of initial states that describes a phase transition
with a spinodal region in field space. 
 
\section{The Model and Equations of Motion}

Having recognized the non-perturbative dynamics of the long wavelength
fluctuations, we need to study the dynamics within a non-perturbative
framework. That is, a framework allowing calculations
for non-perturbatively large energy densities. 
We require that such a framework be: i) renormalizable, ii)
covariant energy conserving, iii) numerically implementable.  There are very
few schemes that fulfill all of these criteria: the Hartree and the large $ N $
approximation\cite{vilenkin,us1,De Sitter}. Whereas the Hartree approximation
is basically a Gaussian variational approximation\cite{jackiwetal,guven} that
in general cannot be consistently improved upon, the large $ N $ approximation
can be consistently implemented beyond leading order\cite{motola,largen} and in
our case it has the added bonus of providing many light fields (associated with
Goldstone modes) that will permit the study of the effects of other fields
which are lighter than the inflaton on the dynamics. Thus we will study the
inflationary dynamics of a quenched phase transition within the framework of
the large $ N $ limit of a scalar theory in the vector representation of $ O(N)
$.

We assume that the universe is spatially flat with a metric given by
\begin{equation}
ds^2 = dt^2 - a^2(t)\, d\vec{x}^2 \; . \label{FRW}
\end{equation}
The matter action and Lagrangian density are given by
\begin{equation}
S_m  =  \int d^4x\; {\cal L}_m = \int d^4x \;
a^3(t)\left[\frac{1}{2}\dot{\vec{\Phi}}^2(x)-\frac{1}{2} 
\frac{(\vec{\nabla}\vec{\Phi}(x))^2}{a^2(t)}-V(\vec{\Phi}(x))\right]
\label{action}
\end{equation}
\begin{equation}
V(\vec{\Phi})  =  \frac{\lambda}{8N}\left(\vec{\Phi}^2-\frac{2N
m^2}{\lambda}\right)^2 +\frac12 \, \xi\; {\cal R} \;\vec{\Phi}^2 
\;, \label{potential}
\end{equation}
\begin{equation}
{\cal R}(t)  =  6\left(\frac{\ddot{a}(t)}{a(t)}+
\frac{\dot{a}^2(t)}{a^2(t)}\right), \label{ricciscalar}
\end{equation}
where we have included the coupling of $\Phi(x)$ to the scalar curvature ${\cal
R}(t)$ since it will arise as a consequence of renormalization\cite{frwpaper}. 

The gravitational sector includes the usual Einstein term in addition
to a higher order curvature term and a cosmological constant term 
which are necessary to renormalize the theory. The action for
the gravitational sector is therefore:
\begin{equation}
S_g  =  \int d^4x\; {\cal L}_g = \int d^4x \, a^3(t) \left[\frac{{\cal
R}(t)}{16\pi G}  
+ \frac{\alpha}{2}\; {\cal R}^2(t) - K\right].
\end{equation}
with K being the cosmological constant (we use $K$ rather than the conventional
$\Lambda/8\pi G$ to distinguish the cosmological constant from the 
ultraviolet cutoff $\Lambda$ we introduce to regularize the theory;
see section IV).
In principle, we also need to include the terms $R^{\mu\nu}R_{\mu\nu}$
and $R^{\alpha\beta\mu\nu}R_{\alpha\beta\mu\nu}$ as they are also terms
of fourth order in derivatives of the metric (fourth adiabatic order),
but the variations resulting from these terms turn out not to be 
independent of that of ${\cal R}^2$ in the flat
FRW cosmology we are considering.

The variation of the action $S = S_g + S_m$ with respect to the 
metric $g_{\mu\nu}$ gives us Einstein's equation
\begin{equation}
\frac{G_{\mu\nu}}{8\pi G} + \alpha H_{\mu\nu} + K g_{\mu\nu}
= - <T_{\mu\nu}>\; ,
\label{extendEinstein}
\end{equation}
where $G_{\mu\nu}$ is the Einstein tensor given by the variation of
$\sqrt{-g}{\cal R}$, $H_{\mu\nu}$ is the higher order curvature term given
by the variation of $\sqrt{-g}{\cal R}^2$, and $T_{\mu\nu}$ is the contribution 
from the matter Lagrangian. 
 With the metric (\ref{FRW}), the various components
of the curvature tensors in terms of the scale factor are:
\begin{eqnarray}
G^{0}_{0} & = & -3(\dot{a}/a)^2, \\
G^{\mu}_{\mu} & = & -{\cal R} = -6\left(\frac{\ddot{a}}{a}
+\frac{\dot{a}^2}{a^2}\right), \\ 
H^{0}_{0} & = & -6\left(\frac{\dot{a}}{a}\dot{{\cal R}} + 
\frac{\dot{a}^2}{a^2}{\cal R} - \frac{1}{12}{\cal R}^2\right), 
\label{hzerozero} \\
H^{\mu}_{\mu} & = & -6\left(\ddot{{\cal R}} + 
3\frac{\dot{a}}{a}\dot{{\cal R}}\right).
\label{htrace}
\end{eqnarray}
Eventually, when we have fully renormalized the theory,
we will set $\alpha_R=0$ and keep as our only contribution to
$K_R$ a piece related to the matter fields which we shall 
incorporate into $T_{\mu\nu}$.  

\subsection{\bf The Large $N$ Approximation} 
To obtain the proper large $N$ limit, the vector field is written as
$$
\vec{\Phi}(\vec x, t) = (\sigma(\vec x,t), \vec{\pi}(\vec x,t)),
$$ 
with $\vec{\pi}$ an $(N-1)$-plet, and we write
\begin{equation}
\sigma(\vec x,t) = \sqrt{N}\phi(t) + \chi(\vec x,t) \; \; ; \; \; \langle
\sigma(\vec x, t) \rangle= \sqrt{N}\phi(t) \; \; ; \; \; 
\langle \chi(\vec x,t) \rangle = 0.
\label{largenzeromode} 
\end{equation}
To implement the large $N$ limit in a consistent manner, one may introduce an
auxiliary field as in\cite{largen}.  However, the leading order
contribution can be obtained equivalently by invoking the 
factorization\cite{De Sitter,frw2}:

\begin{eqnarray}
\chi^4 & \rightarrow & 6 \langle \chi^2 \rangle \chi^2 +\mbox{ constant },
\label{larg1} \\ \chi^3 & \rightarrow & 3 \langle \chi^2 \rangle \chi,
\label{larg2} \\ \left( \vec{\pi} \cdot \vec{\pi} \right)^2 & \rightarrow &
2 \langle \vec{\pi}^2 \rangle \vec{\pi}^2 - \langle \vec{\pi}^2 \rangle^2+
{\cal{O}}(1/N), \label{larg3} \\ \vec{\pi}^2 \chi^2 & \rightarrow & \langle
\vec{\pi}^2 \rangle \chi^2 +\vec{\pi}^2 \langle \chi^2 \rangle,
\label{larg4} \\ \vec{\pi}^2 \chi & \rightarrow & \langle \vec{\pi}^2
\rangle \chi.  \label{larg5}
\end{eqnarray}

To obtain a large $N$ limit, we define\cite{De Sitter,frw2}
\begin{equation} 
\vec{\pi}(\vec x, t) = \psi(\vec x, t)
\overbrace{\left(1,1,\cdots,1\right)}^{N-1}, \label{filargeN}
\end{equation} 
where the large $N$ limit is implemented by the requirement that
\begin{equation}
\langle \psi^2 \rangle \approx {\cal{O}} (1) \; , \; \langle \chi^2 \rangle
\approx {\cal{O}} (1) \; , \; \phi \approx {\cal{O}} (1).
\label{order1}
\end{equation}
The leading contribution is obtained by neglecting the $ {\cal{O}} ({1}\slash
{N})$ terms in the formal limit. The resulting Lagrangian density is
quadratic, with  linear terms in $\chi$ and $\vec{\pi}$.  
The equations of motion are obtained by imposing the tadpole conditions
$<\chi(\vec x,t)>=0$ and $<\vec{\pi}(\vec x,t)> =0$ which in this case are
tantamount to requiring that the linear terms in $\chi$ and $\vec{\pi}$ in
the Lagrangian density vanish. 
Since the action is quadratic, the quantum fields can be expanded in terms
of creation and annihilation operators and mode functions that obey the
Heisenberg equations of motion
\begin{equation} 
\vec{\pi}(\vec x, t) = \int\frac{d^3k}{(2\pi)^3}
\left[{\vec a}_k \; f_k(t) \; e^{i\vec{k}\cdot \vec x} + {\vec a}^{\dagger}_k
 \; f^*_k(t) \; e^{-i\vec{k}\cdot \vec x} \right] .
\end{equation}
The tadpole condition leads to the following equations
of motion\cite{De Sitter,frw2}:

\begin{equation}
\ddot{\phi}(t)+3H(t) \; \dot{\phi}(t)+{\cal M}^2(t) \; \phi(t)=0,
\label{largezeromodeeqn} 
\end{equation}
with the mode functions
\begin{equation}
\left[\frac{d^2}{dt^2}+3H(t)\frac{d}{dt}+\frac{k^2}{a^2(t)}+{\cal
M}^2(t) \right]f_k(t)= 0, 
\label{largenmodes}
\end{equation}
where
\begin{equation}
{\cal M}^2(t) =  -m^2+\xi{\cal R}+ \frac{\lambda}{2}\phi^2(t)+
\frac{\lambda}{2}\langle \psi^2(t) \rangle \; .
\label{Ngranmass}
\end{equation}

In this leading order in $ 1/N $ the theory becomes Gaussian, but with the
self-consistency condition
\begin{equation}
\langle \psi^2(t) \rangle = \int
\frac{d^3k}{(2\pi)^3}\frac{|f_k(t)|^2}{2}.
\label{largenfluc}
\end{equation}

The initial conditions on the modes $f_k(t)$ must now be determined. 
At this stage it proves illuminating to pass to conformal time variables
in terms of the conformally rescaled fields (see \cite{frw2} and
section IX for a discussion)
in which the mode functions obey an equation which is very similar to that
of harmonic oscillators with time dependent frequencies in Minkowski
space-time.  
It has been realized that different initial conditions on the mode functions
lead to different renormalization counterterms\cite{frw2};
 in particular imposing initial conditions in comoving time leads to
counterterms that depend on these initial conditions. Thus we chose to
impose initial conditions in conformal time in terms of the conformally
rescaled mode functions leading to the following initial conditions
in comoving time:
\begin{equation}
f_k(t_0)=\frac{1}{\sqrt{W_k}}, \;\;\; 
\dot{f}_k(t_0)=\left[-\frac{\dot{a}(t_0)}{a(t_0)}-iW_k\right]f_k(t_0),
\label{initcond}
\end{equation}
with
\begin{equation}
W_k^2 \equiv k^2 + {\cal M}^2(t_0) - \frac{{\cal R}(t_0)}{6}.
\label{freq}
\end{equation}
For convenience, we have set $a(t_0)=1$ in eq. (\ref{freq}).
At this point we recognize that when ${\cal M}^2(t_0) - {\cal R}(t_0)/6 <0$ 
the above initial condition must be modified to avoid imaginary frequencies,
which are the signal of instabilities for long wavelength modes. Thus
we {\em define} the initial frequencies that determine the initial conditions
(\ref{initcond}) as
\begin{eqnarray}
W_k^2 & \equiv &  k^2 + \left|{\cal M}^2(t_0) - \frac{{\cal
R}(t_0)}{6}\right| \; \mbox{ for } 
k^2 < \left |{\cal M}^2(t_0) - \frac{{\cal R}(t_0)}{6} \right|\; ,
\label{unstcond1} \\ 
W_k^2 & \equiv &  k^2 + {\cal M}^2(t_0) - \frac{{\cal R}(t_0)}{6} \; \mbox{ for }
k^2 \geq \left |{\cal M}^2(t_0) - \frac{{\cal R}(t_0)}{6} \right|\; . \label{unstcond2}
\end{eqnarray}
As an alternative we have also used initial conditions which smoothly interpolate
from positive frequencies for the unstable modes to the adiabatic vacuum initial
conditions defined by (\ref{initcond},\ref{freq}) for the high $k$ modes.
While the alternative choices of initial conditions result in small quantitative
differences in the results (a few percent in quantities which depend strongly
on these low-$k$ modes), all of the qualitative features we will examine
are independent of this choice.

In the large $N$ limit we find the energy density and pressure 
density to be given by\cite{De Sitter,frw2}
\begin{eqnarray}
\frac{\varepsilon}{N} & = & \frac12\dot{\phi}^2 + \frac12m^2\phi^2 +
\frac{\lambda}{8}\phi^4 + \frac{m^4}{2\lambda} - \xi\;
G^{0}_{0}\;\phi^2 +  6\xi\;\frac{\dot{a}}{a}\;\phi\;\dot{\phi}
\nonumber \\
& + & \frac12\langle\dot{\psi}^2\rangle + 
\frac{1}{2a^2}\langle(\nabla\psi)^2\rangle + \frac12 m^2 
\langle\psi^2\rangle 
+ \frac{\lambda}{8}[2\phi^2\;\langle\psi^2\rangle + \langle\psi^2\rangle^2] 
\nonumber \\
& - & \xi G^{0}_{0}\langle\psi^2\rangle + 
6\xi\;\frac{\dot{a}}{a}\;\langle\psi\dot{\psi}\rangle, \label{energy} \\
\frac{\varepsilon-3p}{N} & = & -\dot{\phi}^2 + 2m^2\;\phi^2 + 
\frac{\lambda}{2}\;\phi^4 + \frac{2m^4}{\lambda} - \xi
G^{\mu}_{\mu}\phi^2 + 
6\xi\left(\phi\ddot{\phi} + \dot{\phi}^2 +
3\frac{\dot{a}}{a}\phi\dot{\phi}\right) 
\nonumber \\
& - & \langle\dot{\psi}^2\rangle +
\frac{1}{a^2}\langle(\nabla\psi)^2\rangle + 
2m^2\;\langle\psi^2\rangle - \xi\;G^{\mu}_{\mu}\;\langle\psi^2\rangle  
\nonumber \\
& + & \frac{\lambda}{2}[2\phi^2\;\langle\psi^2\rangle +
\langle\psi^2\rangle^2] + 
6\xi\left(\langle \psi\;\ddot{\psi}\rangle + \langle \dot{\psi}^2 \rangle+
3\,\frac{\dot{a}}{a}\;\langle\psi\dot{\psi}\rangle \right),
\label{trace}
\end{eqnarray}
where $\langle\psi^2\rangle$ is given by equation (\ref{largenfluc}) 
and we have defined the following integrals:
\begin{eqnarray}
\langle(\nabla\psi)^2\rangle & = & \int \frac{d^3k}{2 (2\pi)^3}\; k^2\;
|f_k(t)|^2 \; , \label{delpsi} \\
\langle\dot{\psi}^2\rangle & = & \int \frac{d^3k}{2(2\pi)^3}\;
|\dot{f}_k(t)|^2 \;. \label{dotpsi}
\end{eqnarray}
The composite operators $\langle \psi \dot{\psi} \rangle$ and 
$\langle \psi \ddot{\psi} \rangle$ are symmetrized by
removing a normal ordering constant to yield
\begin{eqnarray}
\frac{1}{2}(\langle\psi\dot{\psi}\rangle + \langle\dot{\psi}\psi\rangle)
 &=&  \frac{1}{4} \int \frac{d^3k}{(2\pi)^3}
\frac{d |f_k(t)|^2}{dt}\; , \label{psidotpsi} \\
\frac{1}{2}(\langle\psi\ddot{\psi}\rangle + \langle\ddot{\psi}\psi\rangle)
 &=&  \frac{1}{4} \int \frac{d^3k}{(2\pi)^3}
\left[f_k(t)\ddot{f}_k^*(t)+\ddot{f}_k(t)f_k^*(t)\right] \; . 
\label{psiddotpsi} 
\end{eqnarray}
The last of these integrals, (\ref{psiddotpsi}), may be rewritten using the
equation of motion (\ref{largenmodes}):
\begin{equation}
\langle \psi \ddot{\psi} \rangle = -3\frac{\dot{a}}{a}\langle \;
\psi \dot{\psi} \rangle
-\frac{\langle(\nabla\psi)^2\rangle}{a^2}-{\cal M}^2 \langle\psi^2\rangle \; .
\end{equation}
It is straightforward to show that the bare energy is
covariantly conserved by using the equations of motion for the zero mode and
the mode functions.

\section{Renormalization}

Renormalization is a very subtle but important issue in gravitational
backgrounds\cite{birriel}. The fluctuation contribution
$\langle \psi^2(\vec x,t) \rangle$, the energy, and the pressure all need to be
renormalized. The renormalization aspects in curved space times have been
discussed at length in the literature\cite{birriel} and have
been extended to the large $N$ self-consistent approximations for
the non-equilibrium backreaction problem in\cite{largen,frw2,ramsey}.
More recently a consistent and covariant regularization scheme that can be implemented numerically  has
been provided\cite{baacke}. 

In terms of the effective mass term for the large $ N $ limit given by
(\ref{Ngranmass}) and defining the quantity
\begin{eqnarray}
B(t) &\equiv& a^2(t)\left({\cal M}^2(t)-{\cal{R}}/6 \right),\label{boft}\\
{\cal M}^2(t) &=& -m^2_B+\xi_B {\cal R}(t)+\frac{\lambda_B}{2}\phi^2(t)
+\frac{\lambda_B}{2}\langle\psi^2(t)\rangle_B \; , \label{masso}
\end{eqnarray}
where the subscript $B$ stands for bare quantities,  
we find the following large $k$ behavior for the case of an {\em arbitrary}
scale factor $a(t)$ (with $a(0)=1$):
\begin{eqnarray}
|f_k(t)|^2 &=& \frac{1}{ka^2(t)}- \frac{1}{2k^3
a^2(t)}\;B(t)  \cr \cr &+&
{1 \over {8 k^5 \; a^2(t) }}\left\{  3 B(t)^2 + a(t)
\frac{d}{dt} \left[ a(t) {\dot B}(t) \right]  \right\} +
{\cal{O}}(1/k^7)  \cr \cr
& = & {\cal S}^{(2)}+ {\cal{O}}(1/k^5) \; ,
\label{sub1}\\ 
|\dot{f}_k(t)|^2 &=&
\frac{k}{a^4(t)}+\frac{1}{2ka^4(t)}\left[B(t)+2\dot{a}^2 \right] \cr \cr 
& + & {1 \over {8 k^3 \; a^4(t) }}\left\{ - B(t)^2 - a(t)^2 {\ddot
B}(t) + 3 a(t) 
{\dot a}(t)
{\dot B}(t) - 4 {\dot a}^2(t) B(t) \right\} +  {\cal{O}}(1/k^5) \cr \cr
& = & {\cal S}^{(1)}+ {\cal{O}}(1/k^5) \; ,
\label{sub2}  
\end{eqnarray}
\begin{equation}
\frac12 \left[f_k(t)\dot{f}_k^*(t)+\dot{f}_k(t)f_k^*(t)\right] =
-\frac{1}{k \; a^2(t)}\frac{\dot{a}(t)}{a(t)} - \frac{1}{4 k^3 a^2(t)}
\left[\dot{B}(t) - 2\frac{\dot{a}(t)}{a(t)}\;B(t)\right] +
{\cal{O}}(1/k^5) \; .
\label{sub3}
\end{equation}

Although the divergences can be dealt with by dimensional
regularization, this procedure is not well suited to numerical
analysis (see however ref.\cite{baacke}).  We will make our 
subtractions using an ultraviolet cutoff, $\Lambda a(t)$, constant in {\em physical
coordinates}. This guarantees that the counterterms will be time
independent. The renormalization then proceeds much in the same manner as in
reference\cite{frwpaper}; the quadratic divergences renormalize the mass 
and the logarithmic terms renormalize the quartic coupling and the coupling
to the Ricci scalar. In addition, there is a quartic divergence which
renormalizes the cosmological constant while the leading renormalizations
of Newton's constant and the higher order curvature coupling are quadratic and
logarithmic respectively.
The renormalization conditions on the mass, coupling to the Ricci
scalar and coupling constant are obtained from the requirement that the
frequencies that appear in the mode equations are finite\cite{frwpaper}, i.e:
\begin{equation}
-m^2_B+\xi_B {\cal R}(t)+\frac{ \lambda_B}{2}\phi^2(t)
+\frac{\lambda_B}{2}\langle\psi^2(t)\rangle_B=
-m^2_R+\xi_R {\cal R}(t)+\frac{ \lambda_R}{2}\phi^2(t)
+\frac{\lambda_R}{2}\langle\psi^2(t)\rangle_R, \label{rencond}
\end{equation}
while the renormalizations of Newton's constant, the higher order curvature
coupling, and the cosmological constant are given by the condition of
finiteness of the semi-classical Einstein-Friedmann equation:
\begin{equation}
\frac{G^0_0}{8\pi G_B} + \alpha_B H^0_0 + K_B g^0_0 + 
\langle T^0_0 \rangle_B = \frac{G^0_0}{8\pi G_R} + 
\alpha_R H^0_0 + K_R g^0_0 + \langle T^0_0 \rangle_R \; .
\end{equation}

Finally we arrive at the following set of renormalizations\cite{frw2}:
\begin{eqnarray}
\frac{1}{8\pi N G_R} &=& \frac{1}{8\pi N G_B} 
- 2\left(\xi_R-\frac16\right)\frac{\Lambda^2}{16\pi^2}
+ 2\left(\xi_R-\frac16\right)m_R^2\; \frac{\ln(\Lambda/\kappa)}{16\pi^2}, \\
\frac{\alpha_R}{N} &=& \frac{\alpha_B}{N} 
- \left(\xi_R-\frac16\right)^2\frac{\ln(\Lambda/\kappa)}{16\pi^2}, \\
\frac{K_R}{N} &=& \frac{K_B}{N} - \frac{\Lambda^4}{16\pi^2}
+ m_R^2\;\frac{\Lambda^2}{16\pi^2}
+ \frac{m_R^4}{2}\;\frac{\ln(\Lambda/\kappa)}{16\pi^2}, \\
-m_R^2 &=& -m_B^2 + \lambda_R\;\frac{\Lambda^2}{16\pi^2}
+\lambda_R\; m_R^2\;\frac{\ln(\Lambda/\kappa)}{16\pi^2}, \\
\xi_R &=& \xi_B 
- \lambda_R\left(\xi_R-\frac16\right)\frac{\ln(\Lambda/\kappa)}{16\pi^2}, \\
\lambda_R &=& \lambda_B - \lambda_R\frac{\ln(\Lambda/\kappa)}{16\pi^2},\\
\langle \psi^2(t) \rangle_R &=& \int
\frac{d^3k}{2(2\pi)^3}\left\{|f_k(t)|^2-\left[
\frac{1}{ka^2(t)}- \frac{\Theta(k-\kappa)}{2k^3}
\left[{\cal M}^2(t)-{{{\cal R}(t)}\over 6} \right]
\right] \right\} .
\end{eqnarray}
Here, $\kappa$ is the renormalization point.  As expected, the logarithmic 
terms are consistent with the renormalizations found using dimensional
regularization\cite{baacke,ramsey}.  Again, we set $\alpha_R=0$ and
choose the renormalized cosmological constant such that the vacuum
energy is zero in the true vacuum.  We emphasize that while 
the regulator we have chosen does not respect the covariance of 
the theory, the renormalized energy momentum tensor defined in this 
way nevertheless retains the property of covariant conservation in the
limit when the cutoff is taken to infinity.

The logarithmic subtractions can be neglected because of the coupling
$\lambda \leq 10^{-12}$.  Using the Planck scale as the cutoff and the
inflaton mass $m_R$ as a renormalization point, these terms are of order
$\lambda \ln[M_{pl}/m_R] \leq 10^{-10}$, for $m \geq 10^9 \mbox{ GeV }$. An
equivalent statement is that for these values of the coupling and inflaton
masses, the Landau pole is well beyond the physical cutoff $M_{pl}$.
Our relative 
error in the numerical analysis is of order $10^{-8}$, therefore our numerical
study is insensitive to the logarithmic corrections. Though these corrections
are fundamentally important, numerically they can be neglected. Therefore, in
the numerical computations that follow, we will neglect logarithmic 
renormalization and subtract only
quartic and quadratic divergences in the energy and pressure, and quadratic
divergences in the fluctuation contribution. 

\section{Renormalized Equations of Motion for Dynamical Evolution}

It is convenient to introduce the following dimensionless quantities
and definitions,
\begin{equation}
\tau = m_R t \quad ; \quad h= \frac{H}{m_R} \quad ; 
\quad q=\frac{k}{m_R} \quad \; \quad
\omega_q = \frac{W_k}{m_R} \quad ; \quad g= \frac{\lambda_R}{8\pi^2} \; ,
\label{dimvars1}
\end{equation}
\begin{equation}
\eta^2(\tau) = \frac{\lambda_R}{2m^2_R} \; \phi^2(t)
\quad ; \quad  g\Sigma(\tau) = \frac{\lambda}{2m^2_R}\; \langle \psi^2(t)
\rangle_R  \quad ; \quad f_q(\tau) \equiv \sqrt{m_R} \; f_k(t) \; .
\label{dimvars3}
\end{equation}

Choosing $\xi_R=0$ (minimal coupling)  and the renormalization
 point $\kappa = |m_R|$ and setting $a(0)=1$, 
the equations of motion become:

\vspace{2mm} 

\begin{equation}
\left[\frac{d^2}{d \tau^2}+ 3h \frac{d}{d\tau}-1+\eta^2(\tau)+
g\Sigma(\tau)\right]\eta(\tau) = 0, 
\label{zeromode}
\end{equation}

\begin{eqnarray}
& &\left[\frac{d^2}{d \tau^2}+3h
\frac{d}{d\tau}+\frac{q^2}{a^2(\tau)}-1+\eta^2+g\Sigma(\tau)
\right]f_q(\tau)=0, \nonumber \\ 
& &  f_q(0)  =  \frac{1}{\sqrt{\omega_q}} \quad ; \quad 
\dot{f}_q(0)  = \left[-h(0)-i\omega_q\right]f_q(0), \nonumber \\
& & \omega_q  =  
\left[q^2-1+\eta^2(0)-\frac{{\cal
R}(0)}{6m^2_R}+g\Sigma(0)\right]^{\frac{1}{2}} \; \mbox{ for } \; q^2
> -1+\eta^2(0)-\frac{{\cal R}(0)}{6m^2_R}+g\Sigma(0), \nonumber \\ 
& & \omega_q  =  
\left[q^2+1-\eta^2(0)+\frac{{\cal
R}(0)}{6m^2_R}-g\Sigma(0)\right]^{\frac{1}{2}} \; \mbox{ for } \; q^2
< -1+\eta^2(0)-\frac{{\cal R}(0)}{6m^2_R}+g\Sigma(0) . 
\label{modes}  
\end{eqnarray}
where
$$
\Sigma(\tau)= \int_0^{\infty} q^2 dq \left[ | f_q(\tau)|^2 - {1 \over
{ a(\tau)^2}} + {{\Theta(q - 1)}\over {2 q^3}} \left(/frac{{\cal
M}^2(\tau)}{m^2_R}-{{{\cal{R}(\tau)}}\over{6 m^2_R}}\right)\right] \; .
$$

The initial conditions for $\eta(\tau)$ will be specified later. 
An important point to notice is that the equation of
motion for the $q=0$ mode coincides with that of the zero mode
(\ref{zeromode}). Furthermore, for $\eta(\tau \rightarrow \infty) \neq
0$, a stationary (equilibrium) solution of the eq.(\ref{zeromode})  
is obtained when the sum rule\cite{us1,De Sitter,frw2}
\begin{equation}
-1+\eta^2(\infty)+g\Sigma(\infty) = 0 \label{sumrule}
\end{equation}
is fulfilled. This sum rule is nothing but a consequence of Goldstone's
theorem and is a result of the fact that the large $ N $ approximation 
satisfies the Ward identities associated with the $ O(N) $ symmetry, since
the term  $-1+\eta^2+g\Sigma$ is seen to be the effective mass of the
modes transverse to the symmetry breaking direction, i.e. the Goldstone
modes in the broken symmetry phase.

In terms
of the zero mode $\eta(\tau)$ and the quantum mode function given
by eq.(\ref{modes}) we find that the Friedmann equation for the dynamics
of the scale factor in dimensionless variables is given by

\begin{equation}
 h^2(\tau)    =   4h^2_0 \; \epsilon_R(\tau) \quad  ;   \quad h^2_0 =
 \frac{4\pi N m^2_R}{3M^2_{Pl}\lambda_R} \label{hubblequation} 
 \end{equation}
and the renormalized energy and pressure are given by: 

\begin{eqnarray}
 \epsilon_R(\tau) &  =  &  
\frac{1}{2}\dot{\eta}^2+\frac{1}{4}\left(-1+\eta^2+g\Sigma \right)^2+
 \nonumber \\ 
& &  \frac{g}{2}\int q^2 dq \left[|\dot{f_q}|^2-
{\cal S}^{(1)}(q,\tau)
+\frac{q^2}{a^2}\left(|f_q|^2-\Theta(q - 1)\; {\cal
S}^{(2)}(q,\tau)\right) \right]\; , 
\label{hubble} \\
 (p+\varepsilon)_R(\tau)  & = & \frac{2Nm^4_R}{\lambda_R}\left\{
\dot{\eta}^2 \right. \nonumber \\
& + & \left. g \int q^2 dq \left[|\dot{f_q}|^2-
{\cal S}^{(1)}(q,\tau)
+\frac{q^2}{3a^2}\left(|f_q|^2-\Theta(q - 1)\;{\cal S}^{(2)}(q,\tau)\right)
\right]\right\}, \label{ppluse} 
\end{eqnarray}
where the subtractions ${\cal S}^{(1)}$ and ${\cal S}^{(2)}$ are given
by the right hand sides of eqns.(\ref{sub2}) and (\ref{sub1}) respectively.
 
The renormalized energy and pressure are covariantly conserved:
\begin{equation}\label{concov}
{\dot  \epsilon}_R(\tau) + 3 \, h(\tau)\, (p+\varepsilon)_R(\tau) = 0 \; .
\end{equation}

In order to provide the full
solution we now must provide the values of $\eta(0)$, $\dot{\eta}(0)$,
and $h_0$. Assuming that the 
inflationary epoch is associated with a phase transition at the GUT scale,
this requires that $ N m^4_R/\lambda_R \approx 
(10^{15}\mbox{ Gev })^4 $ and assuming the bound on the scalar
self-coupling $\lambda_R \approx 10^{-12}-10^{-14}$ (this will be seen
later 
to be a compatible requirement), we find that $h_0 \approx N^{1/4}$ which
we will take to be reasonably given by $h_0 \approx 1-10$ (for example
in popular GUT's $ N \approx 20 $ depending on particular representations). 

We will begin by studying the case of most interest from the point of view
of describing the phase transition: $\eta(0)=0$ and $\dot{\eta}(0)=0$,
which are the initial conditions that led to puzzling questions. With
these initial conditions, the evolution equation for the zero mode
eq.(\ref{zeromode}) determines that $\eta(\tau) = 0$ by symmetry.

\subsection{Early time dynamics:}
Before engaging in the numerical study, it proves illuminating to
obtain an estimate of the relevant time scales and an intuitive idea of
the main features of the dynamics. Because the coupling is so weak 
($ g \sim 10^{-12} \ll 1 $) and after
renormalization the contribution from the quantum fluctuations to 
the equations of motion is finite, we can neglect  all the terms
proportional to $ g $ in eqs.(\ref{hubble}) and (\ref{modes}). 

For the case where we choose $\eta(\tau) = 0$ and
the evolution equations for the 
mode functions are those for an inverted oscillator in De Sitter space-time,
which have been studied by Guth and Pi\cite{guthpi}. One
obtains the approximate solution 
\begin{eqnarray}
h(t) & \approx & h_0, \nonumber \\
f_q(t) & \approx & e^{-3 h_0\tau/2} \left[A_q \;
J_{\nu}\left(\frac{q}{h_0}e^{-h_0\tau}\right)+ 
B_q  \; J_{-\nu}\left(\frac{q}{h_0}e^{-h_0\tau}\right)\right], \nonumber \\
\nu & = & \sqrt{\frac{9}{4}+\frac{1}{h^2_0}}, \label{earlytime} 
\end{eqnarray}
where $ J_{\pm \nu}(z) $ are Bessel functions, and $A_q$ and $B_q$ are
determined by the initial conditions on the mode functions:
\begin{eqnarray}
B_q = - \frac{1}{\sqrt{\omega_q}}{{\pi \, q}\over { 2 h_0 \, \sin{\nu \pi} }} 
\; \left[ {{i \omega_q -
\frac12 \, h_0 }\over q} \; J_{\nu}\left(\frac{q}{h_0}\right) - 
J'_{\nu}\left(\frac{q}{h_0}\right) \right],
\label{coefb} \\
A_q =   \frac{1}{\sqrt{\omega_q}}{{\pi \, q}\over { 2 h_0 \, \sin{\nu \pi} }} 
\; \left[ {{i \omega_q -
\frac12 \, h_0 }\over q} \;  J_{-\nu}\left(\frac{q}{h_0}\right) -  
J'_{-\nu}\left(\frac{q}{h_0}\right)\right] \;. \label{coefa}
\end{eqnarray}

After the physical wavevectors cross the horizon, i.e. when $qe^{-h_0
\tau}/h_0 \ll 1$ we find that the mode functions factorize: 
\begin{equation}
f_q(\tau) \approx  {{B_q} \over {\Gamma(1-\nu)}} 
\; \left( {{2h_0\, }\over q}\right)^{\nu}e^{(\nu-3/2)h_0 \tau}. \label{factor}
\end{equation}
 This  result reveals a very
important feature: because of the negative mass squared
term in the matter Lagrangian leading to symmetry breaking (and $\nu > 3/2$), we see
that all of the mode functions {\em grow exponentially} after horizon
crossing (for positive mass squared $ \nu < 3/2 $, and  
they would {\em decrease
exponentially} after horizon crossing). This exponential growth is a 
consequence of the spinodal instabilities which 
is a hallmark of the process of phase separation that
occurs to complete the phase transition. 
 We note, in addition that the time 
dependence is exactly given by that of the $ q=0 $ mode, i.e. the zero 
mode, which is a consequence of the redshifting of the wavevectors and 
the fact that after horizon crossing the contribution of the term
$q^2/a^2(\tau)$ in the equations of motion become negligible.
 We clearly  see that the quantum fluctuations grow exponentially and
they will begin to be of the order of the tree level terms in the
equations of motion when $g\Sigma(\tau) \approx 1$. At large
times 
$$
\Sigma(\tau) \approx {\cal F}^2(h_0) h_0^2
e^{(2\nu-3)h_0 \tau} \; ,  
$$
with
${\cal F}(h_0)$ a finite constant that depends on the initial conditions and
is found numerically to be of ${\cal O}(1)$ [see figure (\ref{fofh})].  

In terms of the initial dimensionful variables, the
 condition  $ g\Sigma(\tau) \approx 1$ translates
to $ <\psi^2(\vec x,t)>_R \approx 2m^2_R/\lambda_R $, i.e. the quantum
fluctuations sample the minima of the (renormalized) tree level potential.
We find that the
time at which the contribution of the 
quantum fluctuations becomes of the same order as the tree level terms is
estimated to be\cite{De Sitter}
\begin{equation}
\tau_s \approx \frac{1}{(2\nu-3)h_0}
\ln\left[\frac{1}{g\, h_0^2 {\cal F}^2(h_0)}\right] 
= \frac32 h_0 \ln\left[\frac{1}{g\, h_0^2 {\cal F}^2(h_0)}\right]
+ {\cal O}(1/h_0).
\label{spinodaltime}
\end{equation}
At this time, the contribution of the quantum fluctuations makes the
back reaction very important and, as will be seen numerically, this
translates into the fact that $\tau_s$ also determines the end of the
De Sitter era and the end of inflation. The total number of e-folds during
the stage of exponential expansion of the scale factor (constant
$h_0$) is  given by   
\begin{equation}
N_e \approx \frac{1}{2\nu-3}\;
\ln\left[\frac{1}{g\; h_0^2\; {\cal F}^2(h_0)}\right] 
= \frac32\; h_0^2\; \ln\left[\frac{1}{g\;  h_0^2 \;{\cal
F}^2(h_0)}\right]  
+ {\cal O}(1)\label{efolds}
\end{equation}
For large $h_0$ we see that the number of e-folds scales as $h^2_0$ as well
as with the logarithm of the inverse coupling. 
These results (\ref{factor},\ref{spinodaltime},\ref{efolds}) will be
confirmed numerically below and will be of paramount importance for the
interpretation of the main consequences of the dynamical evolution. 

\subsection{$\eta(0) \neq 0$: classical or quantum behavior?}

Above we have analyzed the situation when $\eta(0) =0$ (or in dimensionful
variables $\phi(0)=0$). The typical analysis of inflaton dynamics in
the literature involves the {\em classical} evolution of $ \phi(t) $ 
with an initial condition in which $ \phi(0) $ is very close to zero (i.e.
the top of the potential hill) in the `slow-roll' regime, for which
$ \ddot{\phi} \ll 3H\dot{\phi}$. Thus, it is important
to quantify the initial conditions on $ \phi(t) $ for which the
dynamics will be 
determined by the classical evolution of $ \phi(t) $ and those for
which the quantum 
fluctuations dominate the dynamics. We can provide a criterion to
separate classical from quantum dynamics by analyzing the relevant time
scales, estimated by neglecting
non-linearities and backreaction effects. We consider the evolution 
of the zero mode in terms of dimensionless variables, and choose
$ \eta(0) \neq 0 $ and $\dot{\eta}(0) = 0 $.
($\dot{\eta}(0) \neq 0$ simply corresponds
to a shift in origin of time). We  assume  $ \eta(0)^2 << 1
$ which is the relevant case where  spinodal instabilities are important. 
We find
\begin{equation}
\eta(\tau) \approx \eta(0) \; e^{(\nu - \frac{3}{2})h_0\tau}\; .
\end{equation}

The non-linearities will become important and eventually terminate
inflation when $\eta(\tau) \approx 1$. This corresponds to a time scale
given by
\begin{equation}
\tau_c \approx \frac{\ln\left[1/ \eta(0)\right]}{(\nu - \frac{3}{2})\;
h_0}\; .\label{classtime}
\end{equation}
 If $ \tau_c $ is much smaller than the spinodal time $ \tau_s $ given
by eq.(\ref{spinodaltime}) then the {\em classical} evolution of the
zero mode will dominate the dynamics and the quantum fluctuations will
not 
become very large, although they will still undergo spinodal growth. On 
 the other hand, if $\tau_c \gg \tau_s$ the quantum fluctuations will
grow to be very large well before the zero mode reaches the non-linear
regime. In this case the dynamics will be determined completely by
the quantum fluctuations. Then the criterion for the classical or quantum
dynamics is given by
\begin{eqnarray} 
\eta(0) & \gg & \sqrt{g}\;h_0 \Longrightarrow \mbox{ classical dynamics }
\nonumber \\
\eta(0) & \ll & \sqrt{g}\;h_0 \Longrightarrow \mbox{ quantum dynamics }
\label{classquandyn} 
\end{eqnarray}
or in terms of dimensionful variables $\phi(0) \gg H_0$ leads to 
{\em classical dynamics} and $\phi(0) \ll H_0$ leads to 
{\em quantum dynamics}. 

However, even when the classical evolution of the
zero mode dominates the dynamics, the quantum fluctuations grow
exponentially after horizon crossing unless the value of $\phi(t)$ is
very close to the minimum of the tree level potential. In the large $
N $ approximation the spinodal line, that is the values of $\phi(t)$ for  
which there are spinodal instabilities, reaches all the way to the minimum
of the tree level potential as can be seen from the equations of motion for
the mode functions. 
Therefore even in the
classical case one must understand how to deal with quantum fluctuations
that grow after horizon crossing.  

\subsection{Numerics}
\begin{figure}
\epsfig{file=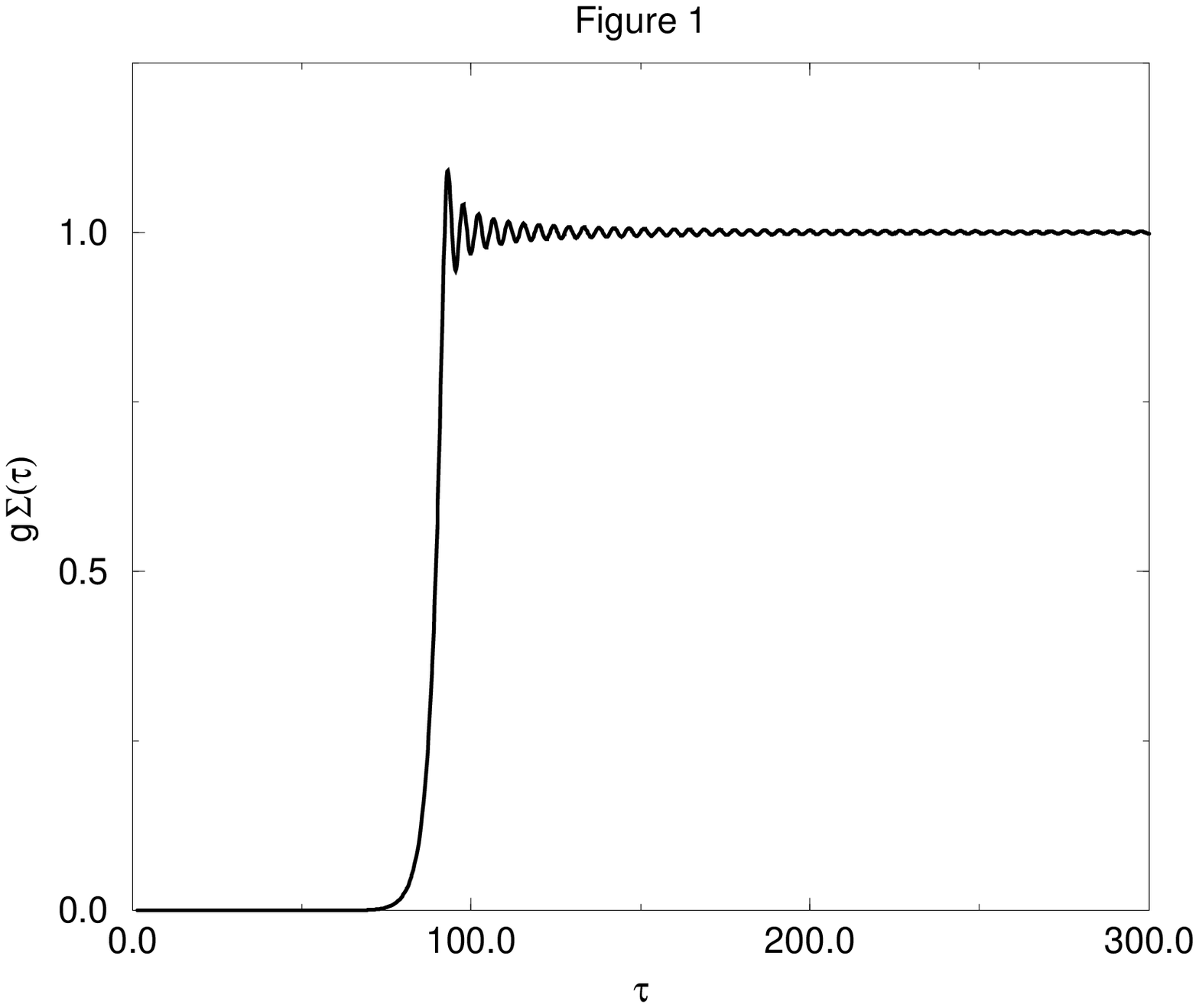,width=5.5in,height=3.2in}
\caption{ $g\Sigma$ vs. $\tau$, for $\eta(0)=0, \dot{\eta}(0)=0,
g = 10^{-14}, h_0 = 2.0$. }
\label{gsigma}
\end{figure}

\begin{figure}
\epsfig{file=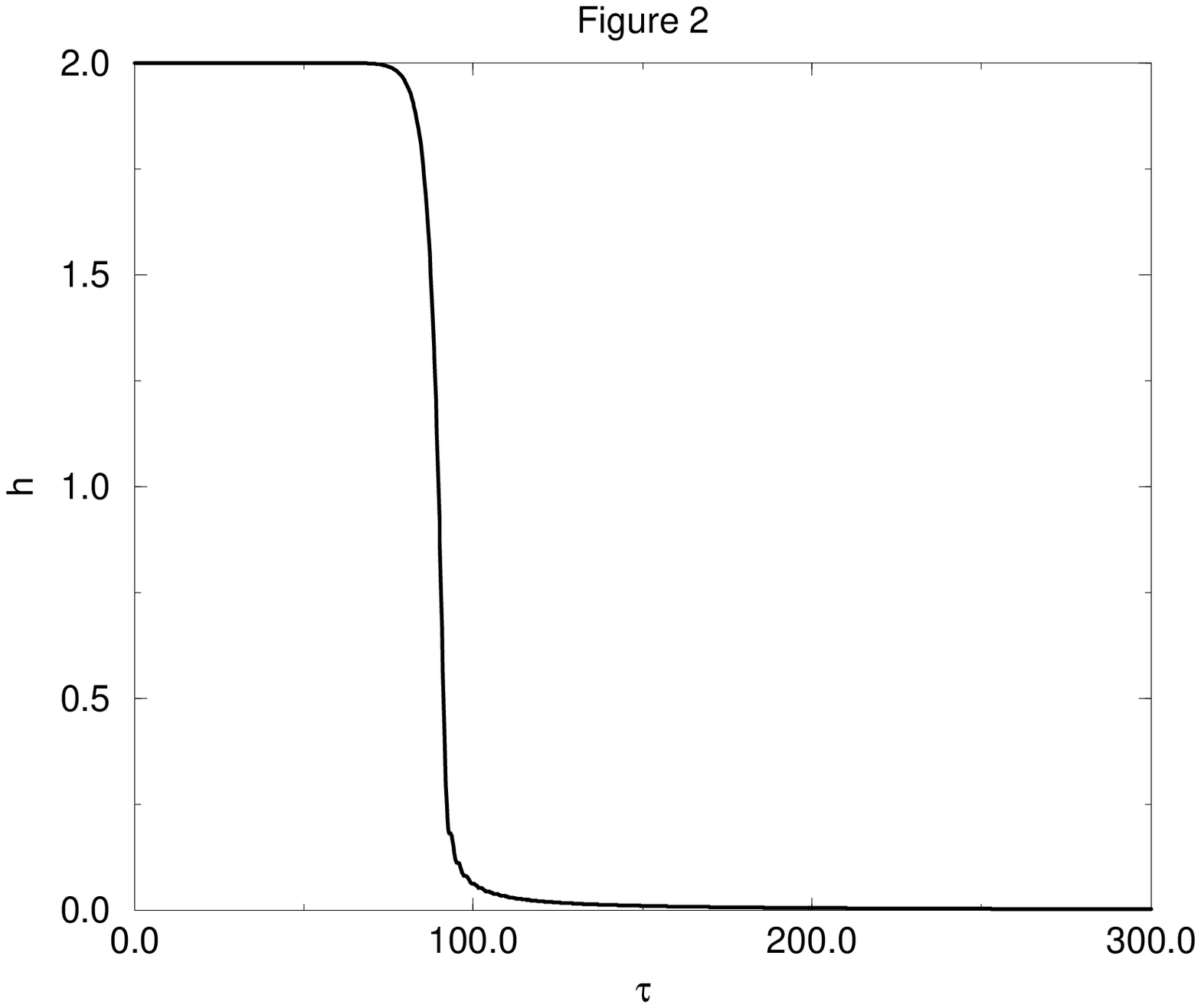,width=5.5in,height=3.2in}
\caption{$H(\tau)$ vs. $\tau$, for $\eta(0)=0, \dot{\eta}(0)=0, 
g = 10^{-14}, h_0 = 2.0 $. }
\label{hubblefig}
\end{figure}

\begin{figure}
\epsfig{file=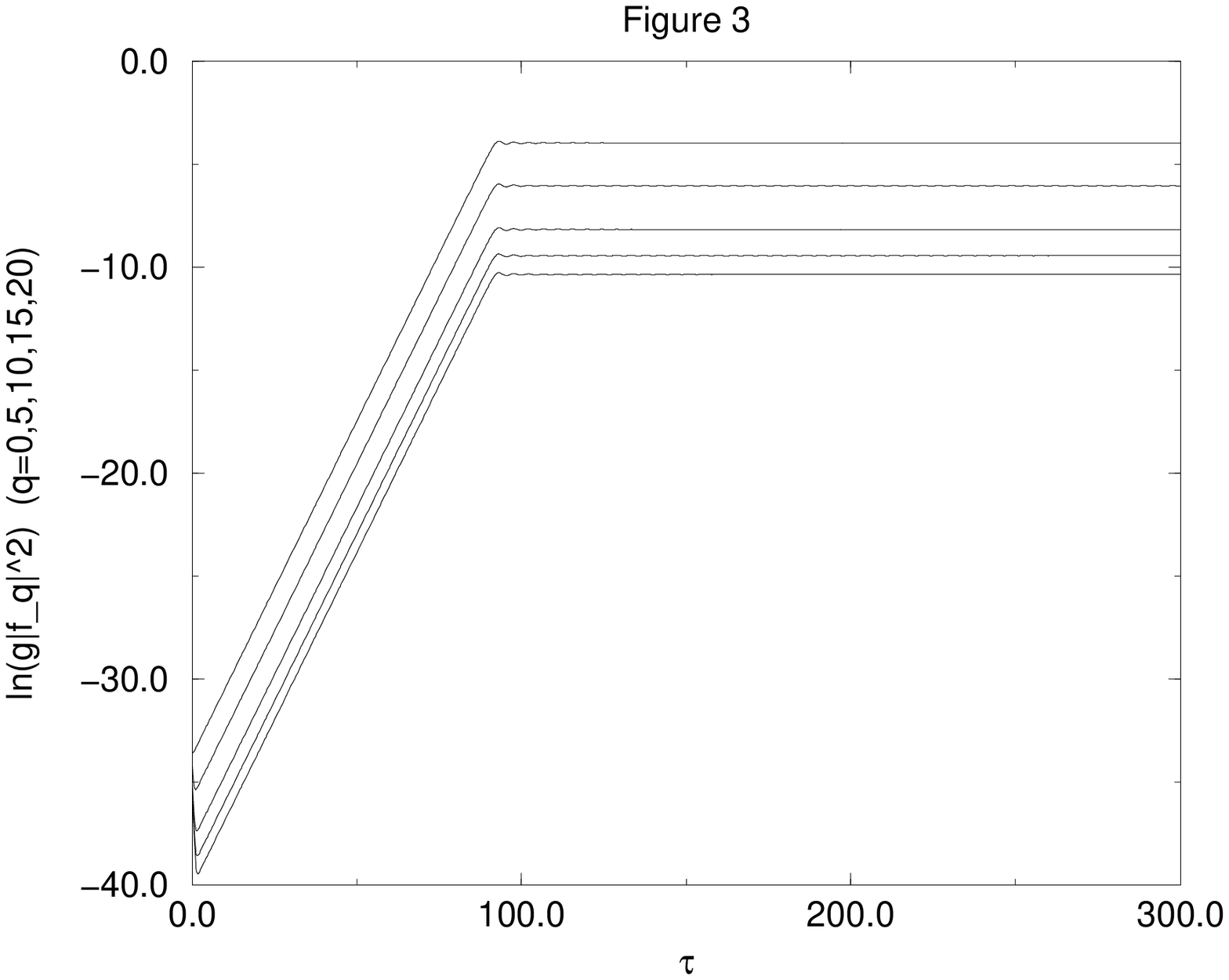,width=5.5in,height=3.2in}
\caption{$\ln(|f_q(\tau)|^2)$ vs. $\tau$, for $\eta(0)=0,
\dot{\eta}(0)=0,  g = 10^{-14}, h_0=2.0$ for
$q=0.0,5,10,15,20$ with smaller 
$q$ corresponding to larger values of $\ln(|f_q(\tau)|^2)$.}
\label{modu}
\end{figure}

\begin{figure}
\epsfig{file=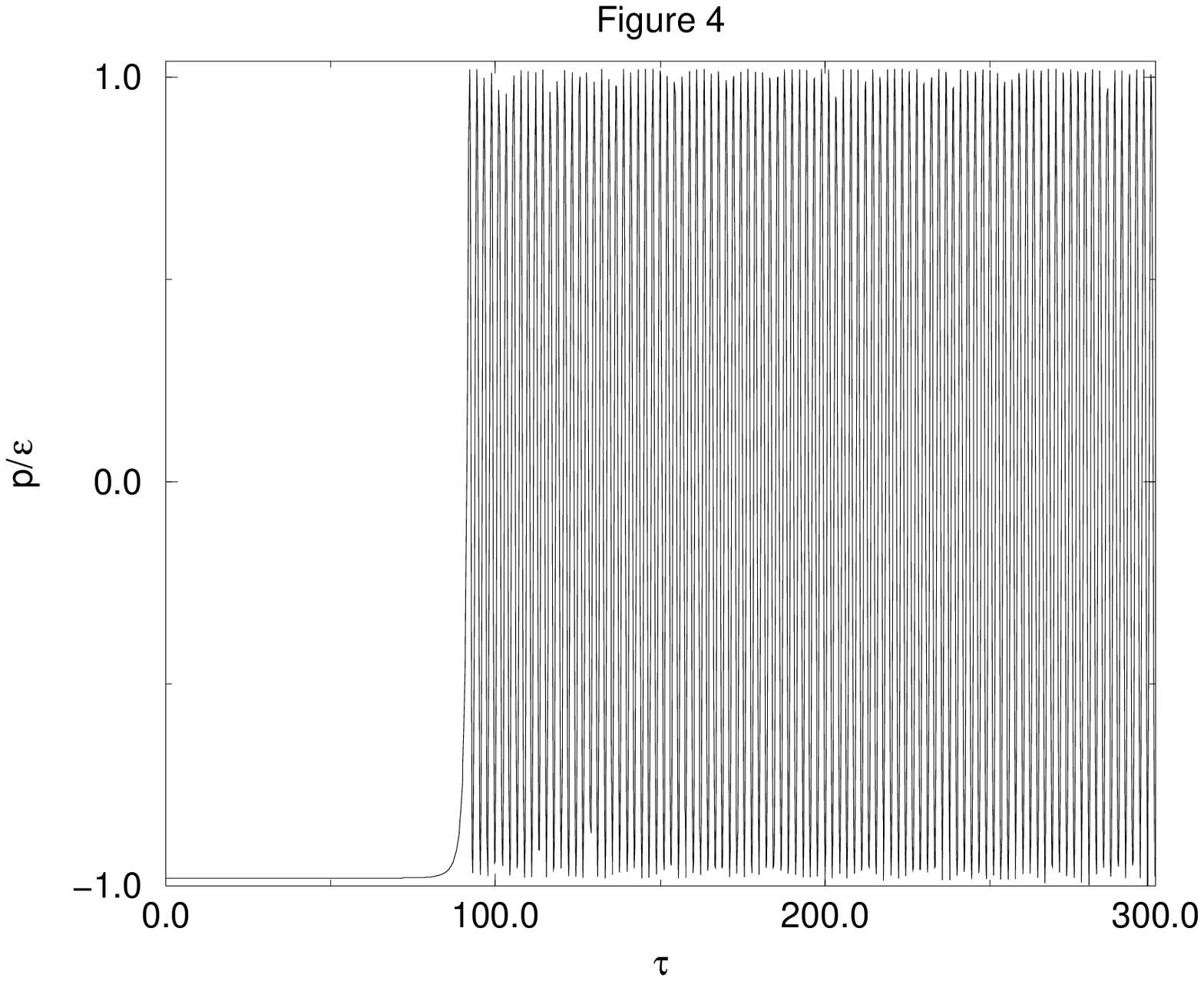,width=5.5in,height=3.2in}
\caption{$p/\varepsilon$ vs. $\tau$, for $\eta(0)=0, \dot{\eta}(0)=0,
g = 10^{-14}, h_0=2.0$.}
\label{povere}
\end{figure}

\begin{figure}
\epsfig{file=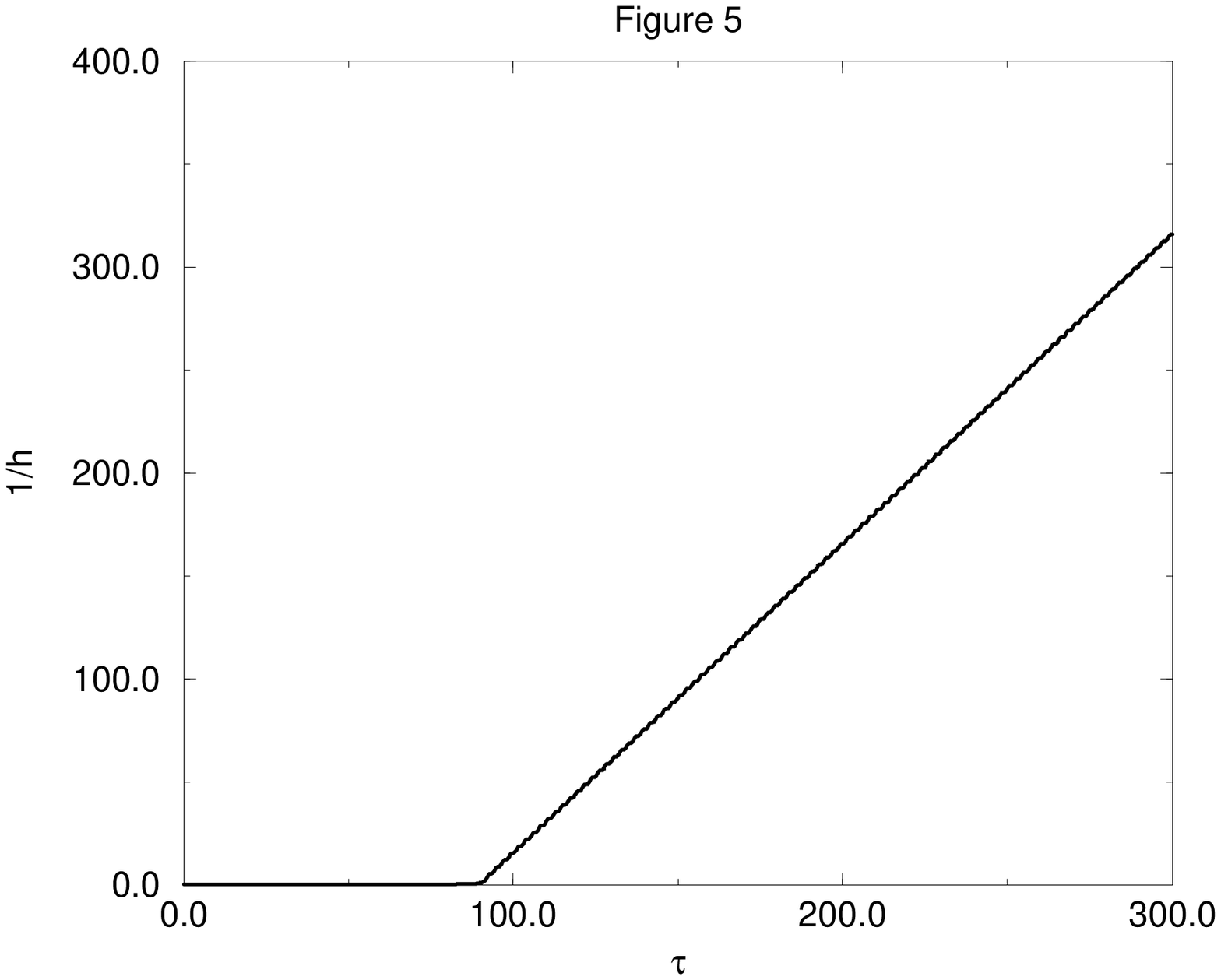,width=5.5in,height=3.2in}
\caption{$1/h(\tau)$ vs. $\tau$, for $\eta(0)=0, \dot{\eta}(0)=0,
g = 10^{-14}, h_0=2.0$. }
\label{hinverse}
\end{figure}

The time evolution is carried out by means of a fourth order Runge-Kutta
routine with adaptive stepsizing while the momentum 
integrals are carried out using an 11-point Newton-Cotes integrator.  
The relative errors in both
the differential equation and the integration are of order $10^{-8}$.
We find that the energy is covariantly conserved throughout the evolution
to better than a part in a thousand. Figures (\ref{gsigma}--\ref{modu}) 
show $g\Sigma(\tau)$ vs. $\tau$,
$h(\tau)$ vs. $\tau$ and $\ln(|f_q(\tau)|^2)$ vs. $\tau$ for several values
of $q$ with larger $q's$ corresponding to successively lower curves. 
Figures (\ref{povere},\ref{hinverse}) show $p(\tau)/\varepsilon(\tau)$ 
and the horizon size $h^{-1}(\tau)$ for 
$g = 10^{-14} \; ; \; \eta(0)=0 \; ; \; \dot{\eta}(0)=0$
and we have chosen the representative value $h_0=2.0$.

Figures (\ref{gsigma}) and (\ref{hubblefig}) show clearly that 
when the contribution of the quantum
fluctuations $ g\Sigma(\tau) $ becomes of order 1 inflation ends,
and the time scale for $ g\Sigma(\tau) $ to reach ${\cal O}(1)$ is very well
described by  the estimate (\ref{spinodaltime}). From figure 1 we see
that this happens for $\tau =\tau_s\approx 90$, leading to a number of
e-folds  
$N_e \approx 180$ which is correctly estimated by  (\ref{spinodaltime},
\ref{efolds}). 

Figure (\ref{modu}) shows clearly the factorization of the modes after they
cross the horizon as described by eq.(\ref{factor}).
 The slopes of all the curves after they become
straight lines in figure (\ref{modu}) is given exactly by $(2\nu-3)$, whereas the
intercept depends on the initial condition on the mode function and
the larger the value of $ q $ the smaller the intercept because the
amplitude of the mode function is smaller initially. Although the
intercept depends on the initial conditions on the long-wavelength
modes, the slope is independent of the value of $q$ and is the same as
what would be obtained in the linear approximation for the {\em
square} of the zero mode at times 
long enough that the decaying solution can be neglected but short enough
that the effect of the non-linearities is very small.
 Notice from the figure that when inflation ends and
the non-linearities become important all of the modes effectively saturate.
This is also what one would expect from the solution of the zero mode:
exponential growth in early-intermediate times (neglecting the
decaying solution), with a growth exponent
given by $(\nu - 3/2)$ and an asymptotic behavior of small oscillations
around the equilibrium position, which for the zero mode is $\eta =1$, but
for the $q \neq 0$ modes depends on the initial conditions. 
All of the mode functions have this behavior once they cross the horizon.
We have also studied the phases of the mode functions and we found that 
they freeze after horizon crossing in the sense that they become independent
of time. This is natural since both the
real and imaginary parts of $ f_q(\tau) $ obey the same equation but
with different 
boundary conditions. After the physical wavelength crosses the horizon, the
dynamics is insensitive to the value of $q$ for real and imaginary parts and
the phases become independent of time. Again, this is a consequence of the
factorization of the modes.  

The growth of the quantum fluctuations is sufficient to end inflation
at a time given by $\tau_s$ in eq.(\ref{spinodaltime}). Furthermore figure
(\ref{povere}) shows that during the inflationary epoch
$p(\tau)/\varepsilon(\tau)  
\approx -1$ and the end of inflation is rather sharp at $\tau_s$ with
$p(\tau)/\varepsilon(\tau)$ oscillating between $\pm 1$ with zero average
over the cycles, resulting in matter domination. Figure
(\ref{hinverse}) shows this 
feature very clearly; $h(\tau)$ is constant during the De Sitter epoch and
becomes matter dominated after the end of inflation with $h^{-1}(\tau) 
\approx \frac32 (\tau -\tau_s)$. There are small oscillations around
this value because both $p(\tau)$ and $\varepsilon(\tau)$
oscillate. These oscillations 
are a result of small oscillations of the mode functions after they 
saturate, and are also a
feature of the solution for a zero mode. 

\begin{figure}
\epsfig{file=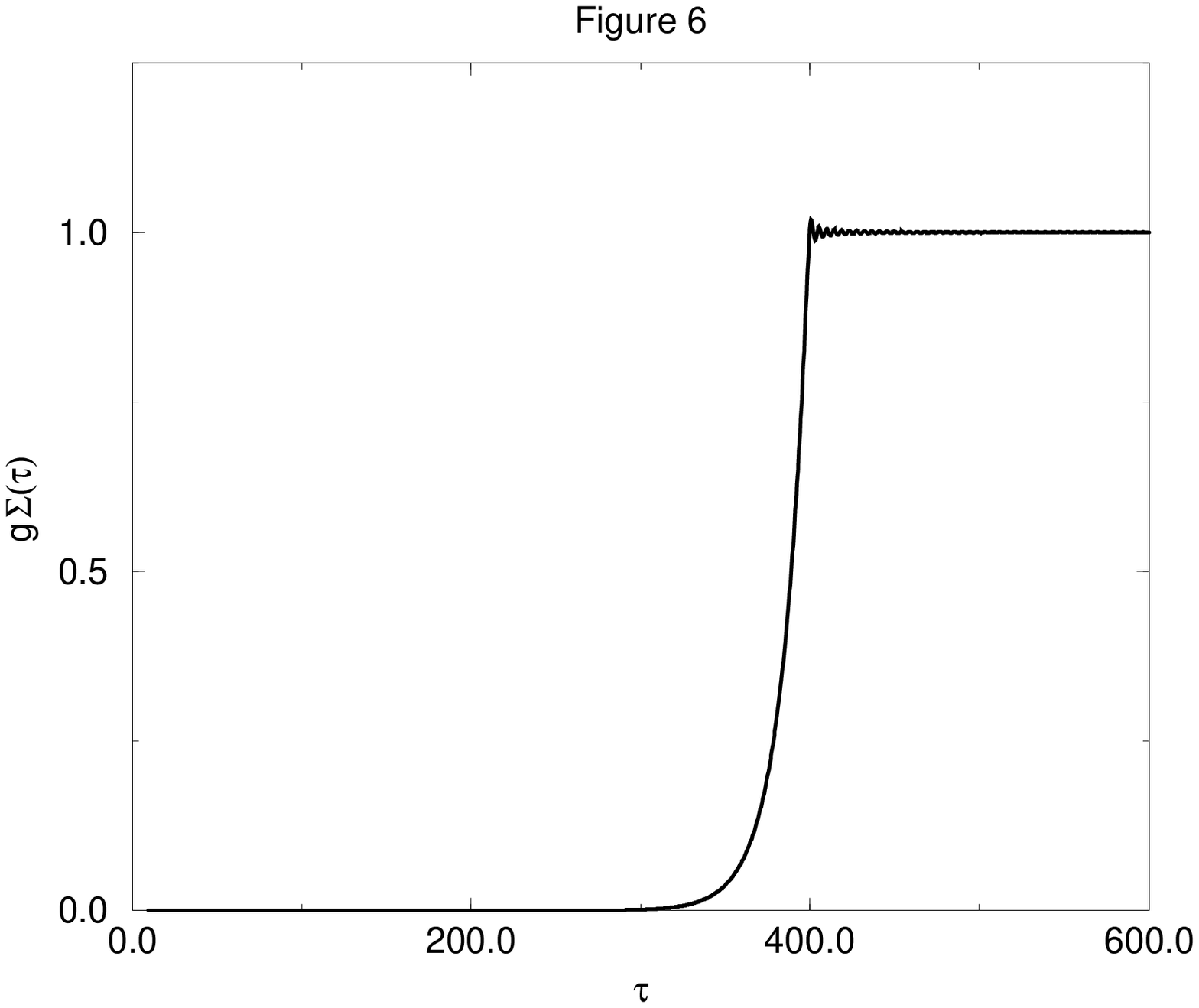,width=5.5in,height=3.2in}
\caption{ $g\Sigma$ vs. $\tau$, for $\eta(0)=0, \dot{\eta}(0)=0,
g = 10^{-14}, h_0 = 10.0$. }
\label{gsigma10}
\end{figure}

\begin{figure}
\epsfig{file=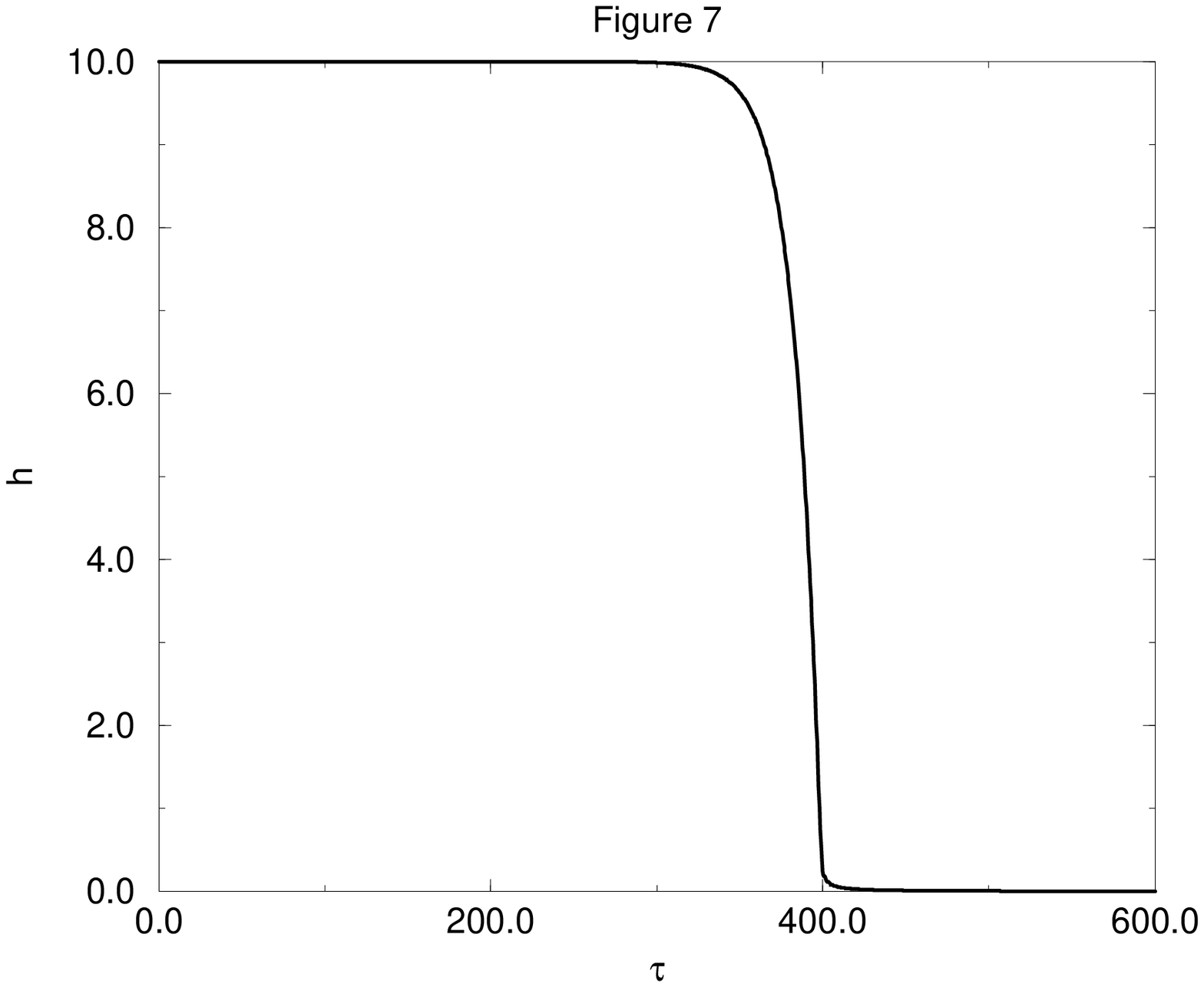,width=5.5in,height=3.2in}
\caption{$H(\tau)$ vs. $\tau$, for $\eta(0)=0, \dot{\eta}(0)=0, 
g = 10^{-14}, h_0 = 10.0 $. }
\label{hubble10}
\end{figure}

\begin{figure}
\epsfig{file=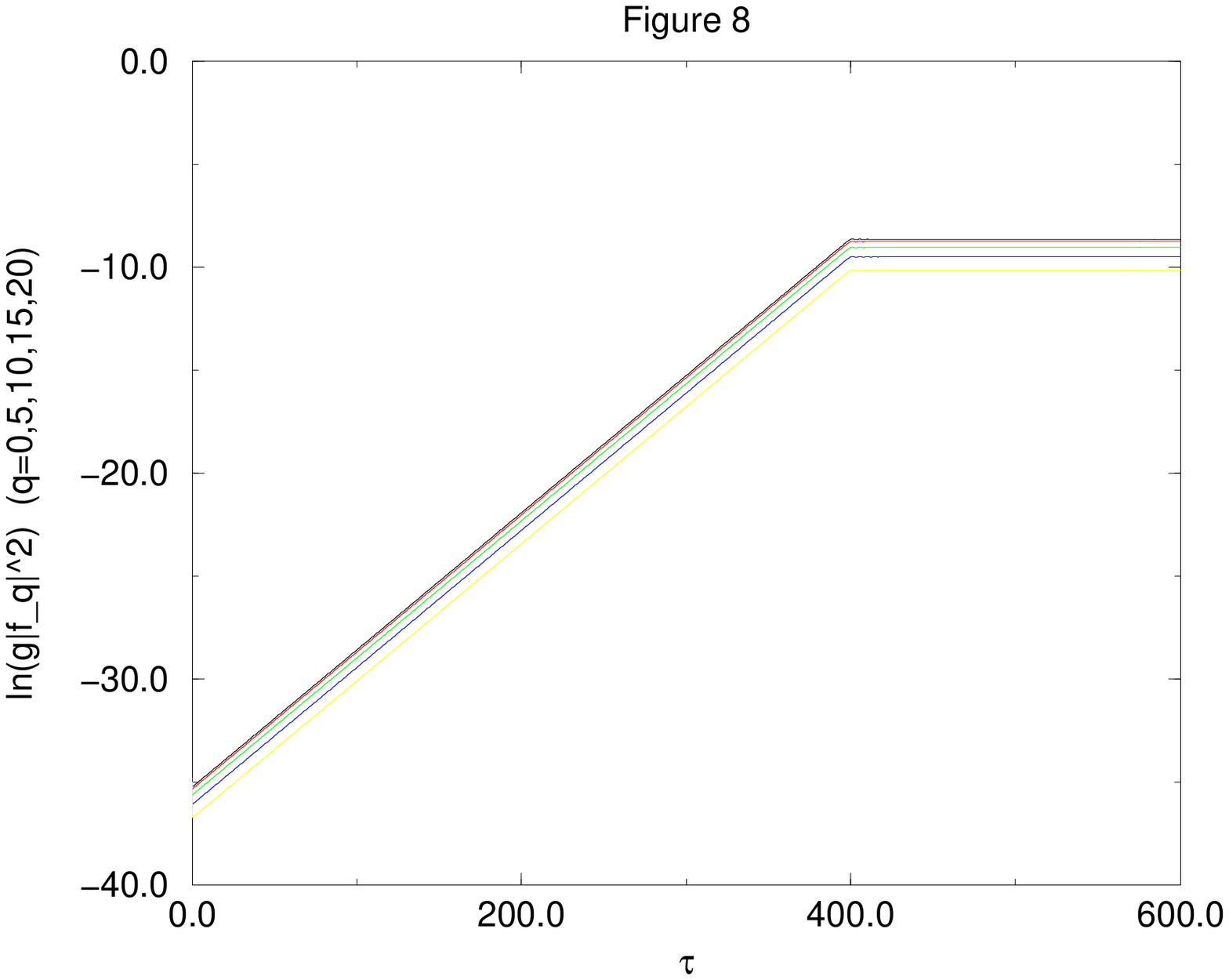,width=5.5in,height=3.2in}
\caption{$\ln(|f_q(\tau)|^2)$ vs. $\tau$, for $\eta(0)=0,
\dot{\eta}(0)=0,  g = 10^{-14}, h_0=10.0$ for
$q=0.0,5,10,15,20$ with smaller 
$q$ corresponding to larger values of $\ln(|f_q(\tau)|^2)$.}
\label{modu10}
\end{figure}

\begin{figure}
\epsfig{file=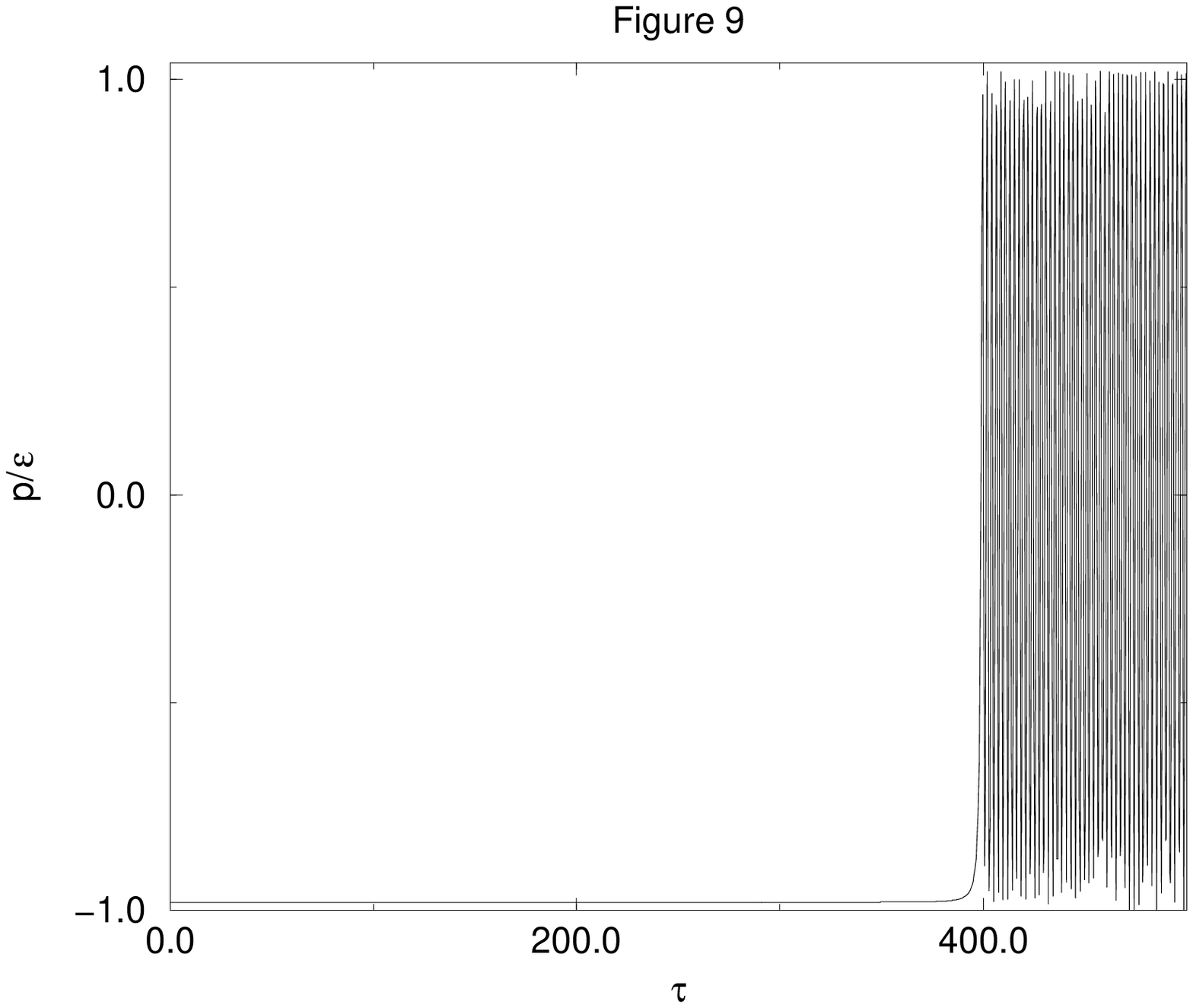,width=5.5in,height=3.2in}
\caption{$p/\varepsilon$ vs. $\tau$, for $\eta(0)=0, \dot{\eta}(0)=0,
g = 10^{-14}, h_0=10.0$.}
\label{povere10}
\end{figure}

\begin{figure}
\epsfig{file=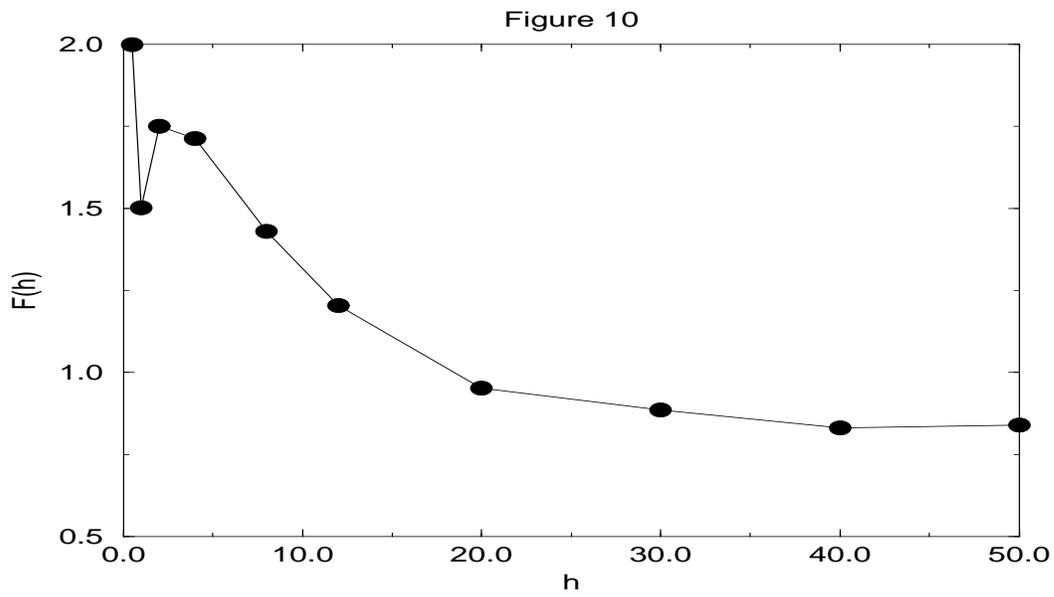,width=5.5in,height=3.2in}
\caption{${\cal F}(H/m)$ vs. $H$, where ${\cal F}(H/m)$
is defined by the relation
$\phi_{eff}(0) = (H/2\pi) {\cal F}(H/m)$ (see eqs.
(\ref{effectivezeromode}) and (\ref{effzeromodein})).}
\label{fofh}
\end{figure}

All of these features hold for a variety of initial conditions.  As an
example, we show in figures (\ref{gsigma10}) -- (\ref{povere10}) the
plots corresponding to figs. (\ref{gsigma}) -- (\ref{povere}) for the
case of an initial Hubble parameter of $h_0=10$.

\subsection{Zero Mode Assembly:}
This remarkable feature of factorization of the mode functions after
horizon crossing can be elegantly summarized as
\begin{equation}
f_k(t)|_{k_{ph}(t) \ll H_0} = g(q,h_0)f_0(\tau),\label{factor2}
\end{equation}
with $k_{ph}(t) = k\,e^{-H_0t}$ being the physical momentum, 
$ g(q,h_0)$ a complex constant, and $f_0(\tau)$ a {\em real} function
of time that satisfies the mode equation with $q=0$ and real initial
conditions which will be inferred later.
 Since the factor $g(q,h_0)$ depends solely on the initial
conditions on the mode functions, it turns out that for two mode
functions corresponding to momenta $k_1,k_2$ that have crossed the
horizon at times $t_1 > t_2$, the ratio of the two mode functions  at
time $t$, ($t_s>t >t_1 >t_2$) is 
$$
{{f_{k_1}(t)}\over {f_{k_2}(t)}} \propto
e^{(\nu-\frac{3}{2})h_0 (\tau_1 - \tau_2)} > 1 \; . 
$$
Then if we consider the 
contribution of these modes to the  {\em renormalized} quantum
fluctuations a long time after the beginning of inflation (so as to
neglect the decaying solutions), we find that 
$$g\Sigma(\tau) \approx {\cal C}e^{(2\nu-3)h_0 \tau} + \mbox{ small} \; , 
$$
where 
`small' stands for
the contribution of mode functions associated with momenta that have not
yet crossed the horizon at time $\tau$, which give a perturbatively
small (of order $\lambda$) contribution.  
 We find that several e-folds
after the beginning of inflation but well before inflation ends, this
factorization of superhorizon modes implies the following:

\begin{eqnarray}
g\int q^2 dq \; |f^2_q(\tau)| & \approx & 
 |C_0|^2 f^2_0(\tau), \label{int1} \\ 
g\int q^2 dq \; |\dot{f}^2_q(\tau)| & \approx & |C_0|^2 \dot{f}^2_0(\tau),
\label{int2} \\
g\int \frac{q^4}{a^2(\tau)} dq \; |f^2_q(\tau)| & \approx & 
 \frac{|C_1|^2 }{a^2(\tau)}f^2_0(\tau), \label{int3}
\end{eqnarray}
where we have neglected the weak time dependence arising from the perturbatively small
contributions of the short-wavelength modes that have not yet crossed the
horizon, and the integrals above are to be understood as the fully
renormalized (subtracted), finite integrals. For $\eta = 0$, we note that (\ref{int1}) and the fact that $f_0(\tau)$ obeys the equation of motion for the mode with $q=0$ leads at once to the conclusion that 
in this regime $\left[g\Sigma(\tau)\right]^{\frac{1}{2}} = |C_0|f_0(\tau)$ obeys the zero mode equation of motion
\begin{equation}
\left[\frac{d^2}{d \tau^2}+ 3h \frac{d}{d\tau}-1+
(|C_0|f_0(\tau))^2\right]|C_0|f_0(\tau) = 0 \; . 
\label{zeromodeeff}
\end{equation}

It is clear that 
several e-folds after the beginning of inflation, we can define an
  effective zero mode   as 
\begin{equation}
\eta^2_{eff}(\tau) \equiv g\Sigma(\tau), \mbox{ or in dimensionful
variables, } \phi_{eff}(t) \equiv \left[\langle \psi^2(\vec x, t)
\rangle_R \right]^{\frac{1}{2}} 
\label{effectivezeromode}
\end{equation}
Although this identification seems natural, we emphasize that it
is by no means a trivial or ad-hoc statement. There are several
important features that allow an {\em unambiguous} identification:
i) $\left[\langle \psi^2(\vec x, t) \rangle_R \right]$ is a fully 
renormalized operator product and hence finite, ii) because of  the
factorization 
 of the superhorizon modes that enter in the evaluation of 
$\left[\langle \psi^2(\vec x, t) \rangle_R \right]$,  
$\phi_{eff}(t)$ (\ref{effectivezeromode}) 
{\em obeys the equation of motion for the zero mode}, iii) this identification 
is valid several e-folds after the beginning of inflation,
after the transient decaying solutions have died away and the integral
in $\langle \psi^2(\vec x,t) \rangle$
is dominated by the modes with wavevector $k$ that have crossed the horizon at 
$t(k) \ll t$.
Numerically we see that this identification holds throughout the
dynamics except for a very few e-folds at the beginning of inflation. This
factorization determines at once the initial conditions of the effective
zero mode that can be extracted numerically: after the first few e-folds and
long before the end of inflation we find
\begin{equation}
\phi_{eff}(t) \equiv \phi_{eff}(0)\; e^{(\nu-
\frac{3}{2})H_0t} \; \; , 
\label{effzeromodein} 
\end{equation}
where we parameterized 
$$
\phi_{eff}(0) \equiv \frac{H_0}{2\pi} \; {\cal F}(H_0/m)
$$
to make contact with the literature. 
As is shown in fig.~(\ref{fofh}), we find numerically that $ {\cal F}(H_0/m)  \approx
{\cal O}(1) $ for a large range of $ 0.1 \leq H_0/m \leq 50 $ and that this quantity
depends on the initial conditions of the long wavelength modes.  

Therefore, in summary, the effective composite zero mode obeys
\begin{equation}
\left[\frac{d^2}{d \tau^2}+ 3h \frac{d}{d\tau}-1+
\eta^2_{eff}(\tau)\right]\eta_{eff}(\tau) = 0 
\; ; \; \dot{\eta}_{eff}(\tau = 0) = (\nu -
\frac{3}{2}) \; \eta_{eff}(0) \; , \label{effzeromode}
\end{equation}
where $ \eta_{eff}(0)
 \equiv {{\sqrt{\lambda_R/2}}\over {m_R}} \; \phi_{eff}(0) $
 is obtained numerically for a given $ h_0 $ by fitting
the intermediate time behavior of  $ g\Sigma(\tau) $ with the growing zero
mode solution.

 Furthermore, this analysis
shows that in the case $\eta = 0$,  the renormalized energy and
pressure in this regime in which the renormalized integrals are
dominated by the superhorizon modes are given by  
\begin{eqnarray}
 \varepsilon_R(\tau) &  \approx   & \frac{2Nm^4_R}{\lambda_R} \left\{
\frac{1}{2}\dot{\eta}^2_{eff}+\frac{1}{4}\left(-1+\eta^2_{eff}\right)^2
\right\},
\label{effenergy} \\
 (p+\varepsilon)_R  & \approx & \frac{2Nm^4_R}{\lambda_R}\left\{
\dot{\eta}^2_{eff}\right\} \label{ppluseeff},
\end{eqnarray}
where we have neglected the contribution proportional to $1/a^2(\tau)$ 
because it is effectively red-shifted away after just a few e-folds.
We found numerically that this term is negligible after the interval
of time necessary for the superhorizon modes to dominate the contribution
to the integrals. 
Then the dynamics of the scale factor is given by 
\begin{equation}
h^2(\tau) = 4 h^2_0 \left\{
\frac{1}{2}\dot{\eta}^2_{eff}+\frac{1}{4}\left(-1+\eta^2_{eff}\right)^2
\right\}.
\label{effscalefactor}
\end{equation}

\begin{figure}
\epsfig{file=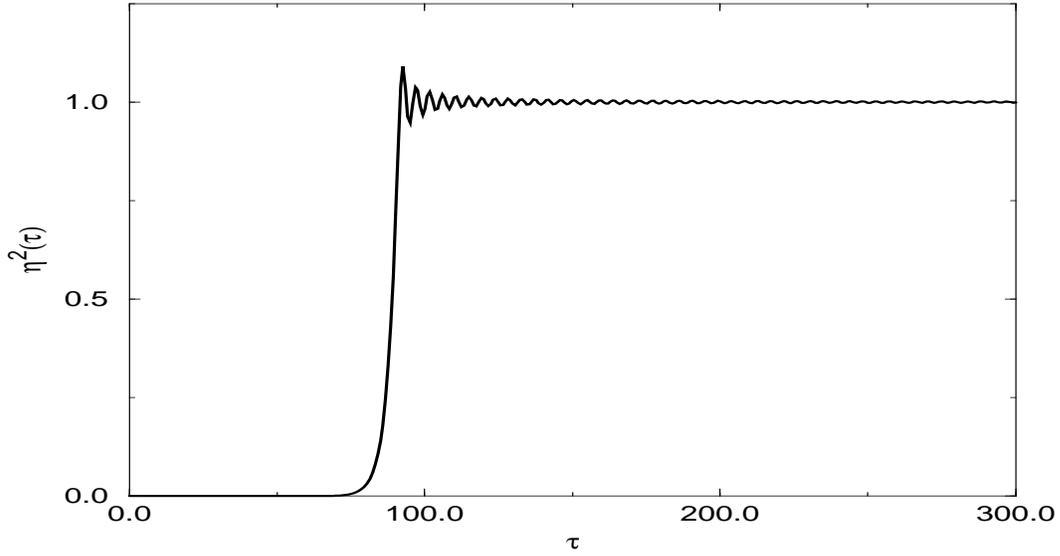,width=5.5in,height=3.2in}
\caption{ $\eta_{eff}^2(\tau)$ vs. $\tau$, for $\eta_{eff}(0)=3.94 \times
10^{-7}, \dot{\eta}_{eff}(0)=0.317\eta_{eff}(0),
g = 10^{-14}, h_0 = 2.0$. The initial conditions were obtained by
fitting the intermediate time regime of $g\Sigma(\tau)$ in
fig. (\ref{gsigma}). $\eta_{eff}(\tau)$ is the solution of
eq.(\ref{effzeromode}) 
with these initial conditions.}
\label{etaclas}
\end{figure}

\begin{figure}
\epsfig{file=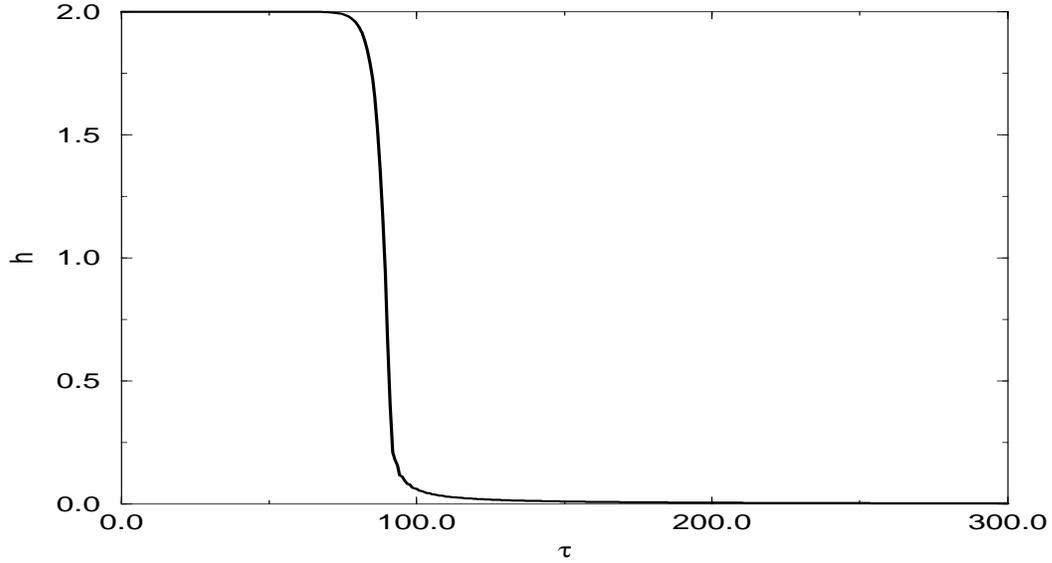,width=5.5in,height=3.2in}
\caption{$h(\tau)$ vs. $\tau$, obtained from the solution of
eqns. (\ref{effzeromode}) and (\ref{effscalefactor}) 
with the conditions of fig. (\ref{etaclas})).}
\label{hubclas}
\end{figure}

We have numerically evolved the set of effective equations (\ref{effzeromode}, \ref{effscalefactor}) by extracting the initial
condition for the effective zero mode from the intermediate time behavior
of $g\Sigma(\tau)$. We found a remarkable agreement  between the
evolution of $\eta^2_{eff}$ and $g\Sigma(\tau)$ and between 
the dynamics of the scale factor in terms of the evolution of
$\eta_{eff}(\tau)$, and the {\em full}
dynamics of the scale factor and quantum fluctuations within our numerical
accuracy. Figures (\ref{etaclas}) and (\ref{hubclas}) show the evolution
of $\eta^2_{eff}(\tau)$ and $h(\tau)$ respectively from the {\em
classical} evolution 
equations (\ref{effzeromode}) and (\ref{effscalefactor}) using the initial
condition  $ \eta_{eff}(0) $ extracted from the exponential fit of
$ g\Sigma(\tau) $ in the intermediate regime. These figures should be 
compared to Figure (\ref{gsigma}) and (\ref{hubblefig}). We have also
numerically compared  
$p/\varepsilon$ given solely by the dynamics of the effective zero mode
and it is again numerically indistinguishable from that obtained with the
full evolution of the mode functions. 

This is one of the main results of our work. In summary: the modes that
become superhorizon sized and grow through the spinodal instabilities assemble
themselves into an effective composite zero mode a few e-folds after
the beginning of inflation. This effective zero mode drives the dynamics
of the FRW scale factor, terminating inflation when the non-linearities
become important. In terms of the underlying fluctuations, the spinodal
growth of superhorizon modes gives a non-perturbatively large contribution
to the energy momentum tensor that drives the dynamics of the scale factor.
Inflation terminates when the mean square root fluctuation probes the
equilibrium minima of the tree level potential. 

This phenomenon of   zero mode assembly, i.e. the `classicalization'
of quantum mechanical fluctuations that grow after horizon crossing is
very similar to the interpretation of `decoherence without decoherence'
of Starobinsky and Polarski\cite{polarski}.

The extension of this analysis to the case for which $\eta(0) \neq 0$
is straightforward.  Since both $\eta(\tau)$ and 
$\sqrt{g\Sigma(\tau)} = |C_0|f_0(\tau)$ obey the equation for the
zero mode, eq.(\ref{zeromode}), it is clear that we can generalize
our definition of the effective zero mode to be
\begin{equation}
\eta_{eff}(\tau) \equiv \sqrt{\eta^2(\tau)+g\Sigma(\tau)}\; .
\label{effeta}
\end{equation}
which obeys the equation of motion of a {\em classical} zero mode:
\begin{equation}
\left[\frac{d^2}{d\tau^2}+3h\frac{d}{d\tau}-1+\eta_{eff}(\tau)^2\right]
\eta_{eff}(\tau) = 0 \; . \label{effzeroeqn}
\end{equation}
If this effective zero mode is to drive the FRW expansion, then the
additional condition
\begin{equation}
\dot{\eta}^2 f_0^2 - 2\eta\dot{\eta}f_0\dot{f_0} + \eta^2 \dot{f_0}^2 = 0
\end{equation}
must also be satisfied.  One can easily show that this relation is indeed
satisfied if the mode functions factorize as in (\ref{factor2}) and if
the integrals (\ref{int1}) -- (\ref{int3}) are dominated by the 
contributions of the superhorizon mode functions.  This leads to the
conclusion that the gravitational dynamics is given by eqns. 
(\ref{effenergy}) -- (\ref{effscalefactor}) with $ \eta_{eff}(\tau) $ defined
by (\ref{effeta}).

We see that in {\em all} cases, the full large $N$ quantum dynamics in these 
models of inflationary phase transitions is well approximated by the
equivalent dynamics of a homogeneous, classical scalar field with initial
conditions on the effective field 
$\eta_{eff}(0) \geq \sqrt{g} h_0 {\cal F}(h_0)$.  
We have verified these
results numerically for the field and scale factor dynamics, finding that
the effective classical dynamics reproduces the results of the full
dynamics to within our numerical accuracy.  
We have also checked numerically
that the estimate for the classical to quantum crossover given by eq.(\ref{classquandyn}) is quantitatively correct. Thus in the classical case in
which $\eta(0) \gg \sqrt{\lambda}\; h_0$ we find that $\eta_{eff}(\tau) = \eta(\tau)$,
 whereas in the opposite, quantum case $\eta_{eff}(\tau) =
\sqrt{g\Sigma(\tau)}$. 

This remarkable feature of zero mode assembly of long-wavelength,
spinodally unstable modes is a consequence of the presence of the horizon.
It also explains why, despite the fact that asymptotically the
fluctuations sample the broken symmetry state, the equation of state is
that of matter. 
Since the excitations in the broken symmetry state are massless Goldstone
bosons one would expect radiation domination. However, the assembly
phenomenon, i.e. the redshifting of the wave vectors, makes these modes behave
exactly like zero momentum modes that give an equation of state of matter
(upon averaging over the small oscillations around the minimum).  

Subhorizon modes at the end of inflation with $ q > h_0 \, e^{h_0
\tau_s} $ do not participate in the zero mode assembly. The behavior of such
modes do depend on $ q $ after the end of inflation. Notice that these
modes have extremely large comoving $ q $ since $  h_0 \, e^{h_0
\tau_s} \geq 10^{26} $. As discussed in ref.\cite{frw2} such modes decrease
with time after inflation as $ \sim 1/a(\tau) $. 

\section{Making sense of small fluctuations:}
Having recognized the effective classical variable that can be interpreted
as the component of the field that drives the FRW background and rolls
down the classical potential hill, we want to recognize unambiguously
the small fluctuations. We have argued above that after horizon crossing,
all of the mode functions evolve proportionally to the zero mode,
and the question arises: which modes are assembled into the effective
zero mode whose dynamics drives the evolution of the FRW scale factor
and which modes are treated as perturbations? In principle every
$k\neq 0$ mode provides some spatial inhomogeneity, and assembling these
into an effective homogeneous zero mode seems in principle to do away with
the very inhomogeneities that one wants to study. However, scales of cosmological importance today first crossed the horizon during the
last 60 or so e-folds of inflation. Recently Grishchuk\cite{grishchuk}
 has argued that the
sensitivity of the measurements of $ \Delta T/T $ probe inhomogeneities on
scales $\approx 500$ times the size of the present horizon. Therefore scales
that are larger than these and that have first crossed the horizon much earlier than the last 60 e-folds of inflation are unobservable today and
can be treated as an effective homogeneous component, whereas the scales that
can be probed experimentally via the CMB inhomogeneities today must be treated 
separately as part of the inhomogeneous perturbations of the CMB. 

Thus a consistent description of the dynamics in terms of an effective
zero mode plus `small' quantum fluctuations can be given provided
the following requirements are met:
a) the total number of e-folds $N_e \gg 60$, b) all the modes that have
crossed the horizon {\em before} the last 60-65 e-folds are assembled into
an effective {\em classical} zero mode via $\phi_{eff}(t) = 
\left[\phi^2_0(t)+ \langle \psi^2(\vec x,t) \rangle_R
\right]^{\frac{1}{2}}$, c) the modes that cross the horizon during the
last 60--65 e-folds are accounted as `small' perturbations. The reason
for the requirement a) is that in the separation 
$\phi(\vec x, t) = \phi_{eff}(t)+\delta \phi(\vec x,t)$ one requires that
$\delta \phi(\vec x,t)/\phi_{eff}(t) \ll 1$. As argued above, after the 
modes cross the horizon, the ratio of amplitudes of the mode functions remains
constant and given by $e^{(\nu - \frac{3}{2})\Delta N}$ with $\Delta N$ 
being the number of e-folds between the crossing of the smaller $ k $ and the
crossing of the larger $ k $. Then for $\delta \phi(\vec x, t)$ to be much
smaller than the effective zero mode, it must be that the Fourier components
of $\delta \phi$ correspond to very large $k$'s at the beginning of inflation,
so that the effective zero mode can grow for a long time before the components
of $\delta \phi$ begin to grow under the spinodal instabilities. 
In fact requirement a) is not very severe; in the figures (1-5) we have taken
$h_0 = 2.0$ which is a very moderate value and yet for $\lambda = 10^{-12}$
the inflationary stage lasts for well over 100 e-folds, and as argued above, the
larger $h_0$ for fixed $\lambda$, the longer is the inflationary stage. 
Therefore under this set of conditions, the classical dynamics of the effective zero mode $\phi_{eff}(t)$  drives the FRW background, whereas
the inhomogeneous fluctuations $\delta \phi(\vec x,t)$, which are made up
of Fourier components with wavelengths that are much smaller than the
horizon at the beginning of inflation and that cross the horizon during
the last 60 e-folds, provide the inhomogeneities that seed density
perturbations.

\section{Scalar and Tensor Metric Perturbations:}
\subsection{Scalar Perturbations:}
Having identified the effective zero mode and the `small perturbations',
we are now in position to provide an estimate for the amplitude and spectrum
of scalar metric perturbations. We use the clear formulation by Mukhanov,
Feldman and Brandenberger\cite{mukhanov} in terms of gauge invariant
variables. In particular we focus on the dynamics of the Bardeen
potential\cite{bardeen}, which in longitudinal gauge is identified
with the 
Newtonian potential. The equation of motion for the Fourier components (in
terms of comoving wavevectors) for this variable in terms of the effective
zero mode is\cite{mukhanov}
\begin{equation}
\ddot{\Phi}_k +
\left[H(t)-2\frac{\ddot{\phi}_{eff}(t)}{\dot{\phi}_{eff}(t)}\right] 
\dot{\Phi}_k+\left[\frac{k^2}{a^2(t)}+ 2\left(\dot{H}(t)-H(t)
\frac{\ddot{\phi}_{eff}(t)}{\dot{\phi}_{eff}(t)}\right)\right]\Phi_k =
0 .
\label{bardeen}
\end{equation}

We are interested in determining the dynamics of $\Phi_k$ for those
wavevectors that cross the horizon during the last 60 e-folds before the
end of inflation. During the inflationary stage the numerical analysis
yields  to a very good approximation
\begin{equation}
H(t) \approx H_0  \; ; \; \phi_{eff}(t) = \phi_{eff}(0)\; e^{(\nu-
\frac{3}{2})H_0t}, \label{infla}
\end{equation}
where $H_0$ is the value of the Hubble constant during inflation, leading to 
\begin{equation}
\Phi_k(t) = e^{(\nu -2)H_0t}\left[a_k\;
H^{(1)}_{\beta}\left(\frac{ke^{-H_0t}}{H_0}\right) 
+b_k \; H^{(2)}_{\beta}\left(\frac{ke^{-H_0t}}{H_0}\right)\right]
\; ; \; \beta= \nu-1 \; .
\label{solbardeen}
\end{equation}
The coefficients $a_k,b_k$ are determined by the initial conditions.

Since we are interested in the wavevectors that cross the horizon during
the last 60 e-folds, the consistency for the zero mode assembly and
the interpretation of `small perturbations' requires that there must be
many e-folds before the {\em last} 60. We are then considering wavevectors
that were deep inside the horizon at the onset of inflation. 
 Mukhanov et. al.\cite{mukhanov} show that $\Phi_k(t)$ is related to
the canonical `velocity field' that determines scalar  perturbations
of the metric and which is quantized with Bunch-Davies initial
conditions for the large $k$-mode functions. The relation between
$\Phi_k$ and $v$ and the  
initial conditions on $v$ lead at once to a determination of the
coefficients $a_k$ and $b_k$ for $k >> H_0$\cite{mukhanov} 
\begin{equation}
a_k = -\frac{3}{2} \left[\frac{8\pi}{3M^2_{Pl}}\right] \dot{\phi}_{eff}(0)
\sqrt{\frac{\pi}{2H_0}} \frac{1}{k} \quad ; \quad b_k = 0\; .
\label{coeffs}
\end{equation} 

Thus we find that the amplitude of scalar metric perturbations after 
horizon crossing is given by
\begin{equation}
|\delta_k(t)| = k^{\frac{3}{2}}|\Phi_k(t)| \approx
\frac{3}{2} \left[\frac{8\sqrt{\pi}}{3M^2_{Pl}}\right] \dot{\phi}_{eff}(0)
\left(\frac{2H_0}{k}\right)^{\nu -\frac{3}{2}}
e^{(2\nu-3)H_0t}\; .
\label{perts}
\end{equation}
The power spectrum per logarithmic $k$  interval is given by
$ |\delta_k(t)|^2 $. The time dependence of $ |\delta_k(t)| $ displays the
unstable growth associated with the spinodal instabilities of
super-horizon modes and 
is a hallmark of the phase transition.
This time dependence can be also understood from the constraint equation
that relates the Bardeen potential to the gauge invariant field fluctuations\cite{mukhanov}, which in longitudinal gauge are identified
with $\delta \phi(\vec x,t)$. 
The constraint equation and the evolution equations for the
gauge invariant scalar field fluctuations are\cite{mukhanov}
\begin{equation}
\frac{d}{dt}(a \Phi_k) = \frac{4\pi}{M^2_{Pl}}\;a\; \delta \phi^{gi}_k
\; \dot{\phi}_0\; ,
\label{constraint}
\end{equation}

\begin{equation}
\left[\frac{d^2}{dt^2}+3H \frac{d}{dt}+\frac{k^2}{a^2}+{\cal M}^2 \right]
\delta \phi^{gi}_k-4 \;
\dot{\phi}_{eff}\;\dot{\Phi}_k+2V'(\phi_{eff})\;\Phi_k=0\; . 
\label{gauginv}
\end{equation}

Since the right hand side of (\ref{constraint}) is proportional to
$\dot{\phi}_{eff}/M^2_{Pl} \ll 1 $ during the inflationary epoch in this
model,  we can neglect the terms proportional
to $\dot{\Phi}_k $ and $\Phi_k$ on the left hand side of (\ref{gauginv}),
in which case the equation for the gauge invariant scalar field
fluctuation is the same as for the mode functions. In fact, since $ 
\delta \phi^{gi}_k$ is gauge invariant we can evaluate it in the longitudinal gauge wherein it is identified with the mode functions
$f_k(t)$. Then absorbing a constant of integration in the initial
conditions for the Bardeen variable, we find
\begin{equation}
\Phi_k(t) \approx \frac{4\pi}{M^2_{Pl}a(t)}\int_{t_o}^t
a(t')\;\phi_{eff}(t')\; f_k(t')\; dt' + {\cal O}\left(\frac{1}{M^4_{Pl}}\right) ,
\label{bard}
\end{equation}
and using that $\phi(t) \propto e^{(\nu-3/2)H_0t} $ and that
 after horizon crossing $f_k(t) \propto e^{(\nu-3/2)H_0t}$, one obtains
at once the time dependence of the Bardeen variable after horizon
crossing. In particular the time dependence is found to be $\propto 
e^{(2\nu-3)H_0t}$. It is then clear that the time dependence is a reflection of
the spinodal (unstable) growth of the superhorizon field fluctuations. 

 To obtain the amplitude and spectrum
of density perturbations at {\em second} horizon crossing we use the
conservation law associated with the gauge invariant variable\cite{mukhanov}
\begin{equation}
\xi_k = \frac{2}{3}
\frac{\frac{\dot{\Phi}_k}{H}+\Phi_k}{1+p/\varepsilon} + \Phi_k
\; \; ; \; \; \dot{\xi}_k =0\; , \label{xivar}
\end{equation}
which is valid after horizon crossing of the mode with wavevector $ k $.
Although this conservation law is an exact statement of superhorizon mode
solutions of eq.(\ref{bardeen}), 
we have obtained solutions assuming that during
the inflationary stage $H$ is constant and have neglected the $\dot{H}$ term in
Eq. (\ref{bardeen}). Since during the inflationary stage,
\begin{equation}
\dot{H}(t) = -\frac{4\pi}{M^2_{Pl}}\, \dot{\phi}^2_{eff}(t) \propto
H^2_0 \; \left(\frac{d\eta_{eff}(\tau)}{d\tau}\right)^2\ll H^2_0 \label{Hdot}
\end{equation}
and $\ddot{\phi}/\dot{\phi} \approx H_0$, the above approximation is
justified. We then see that $\phi^2_{eff}(t) \propto
e^{(2\nu-3)H_0t}$ which is the same time dependence as that of
$\Phi_k(t)$. Thus the
term proportional to $1/(1+p/\varepsilon)$ in Eq. 
(\ref{xivar}) is indeed constant in time after horizon crossing. On the other
hand, the term that
does not have this denominator evolves in time but is of order 
$(1+p/\varepsilon) =
-2\dot{H}/3H^2 \ll 1$ with respect to the constant term and therefore can be
neglected. Thus, we confirm that the variable $\xi$ is conserved
up to the small term proportional to $(1+p/\varepsilon)\Phi_k$ which
is negligible during the inflationary stage. 
This small time dependence is consistent with the fact
that we neglected the $\dot{H}$ term in the equation of motion for $\Phi_k(t)$.
 
The validity of the conservation law has been recently studied and confirmed in
different contexts\cite{caldwell,martin}. Notice that we do not have to assume
that $\dot{\Phi}_k$ vanishes, which in fact does not occur.

 However, upon second horizon crossing
it is straightforward to see that $\dot{\Phi}_k(t_f) \approx 0$. The 
reason for this assertion can be seen as follows: eq.(\ref{gauginv}) shows that
at long times, when the effective zero mode is oscillating around the minimum
of the potential with a very small amplitude and when the time dependence of
the fluctuations has saturated (see figure 3), $\Phi_k$ will redshift
as $\approx 1/a(t)$\cite{frw2} and its derivative becomes extremely small. 

 Using
this conservation law,  assuming matter domination at second horizon 
crossing,  and $\dot{\Phi}_k(t_f)\approx 0$\cite{mukhanov}, we find
\begin{equation}
|\delta_k(t_f)| = \frac{12 \, \Gamma(\nu)\,\sqrt{\pi}}{5\, (\nu-\frac{3}{2})\, 
{\cal F}(H_0/m)} \left(\frac{2H_0}{k}\right)^{\nu-\frac{3}{2}},
\label{amplitude}
\end{equation}
where ${\cal F}(H_0/m)$ determines the initial amplitude of the effective
zero mode (\ref{effzeromodein}). 
We can now read the power spectrum per logarithmic $k$ interval
\begin{equation}
{\cal P}_s(k) = |\delta_k|^2 \propto k^{-2(\nu-\frac{3}{2})}.
\end{equation}
leading to the index for scalar density perturbations
\begin{equation}
n_s = 1-2\left(\nu-\frac{3}{2}\right) \; . \label{index}
\end{equation}

For $H_0/m \gg 1$, we can expand $\nu-3/2$ as a series in $m^2/H_0^2$ in
eq. (\ref{amplitude}).  Given that the comoving wavenumber of the mode which 
crosses the horizon $n$ e-folds before the end of inflation is $k=H_0 e^{(N_e-n)}$ 
where $N_e$ is given by (\ref{efolds}), we arrive at the following expression
for the amplitude of fluctuations on the scale corresponding to $n$ 
in terms of the De Sitter Hubble constant and the coupling $\lambda$:
\begin{equation}
|\delta_n(t_f)| \simeq \frac{9 H_0^3}{5\sqrt{2} m^3} \left(2e^n\right)^{m^2/3H_0^2}
\sqrt{\lambda} \left[1+\frac{2m^2}{3H_0^2} \left(\frac76 - \ln 2 - \frac{\gamma}{2}
\right) + {\cal O}\left( \frac{m^4}{H_0^4} \right) \right] \; .
\label{amplitude_n}
\end{equation}
Here, $\gamma$ is Euler's constant.  Note the explicit dependence of the 
amplitude of density perturbations on $\sqrt{\lambda}$.  For $n \approx 60$,
the factor $\exp(nm^2/3H_0^2)$ is ${\cal O}(100)$ for $H_0/m = 2$, while
it is ${\cal O}(1)$ for $H_0/m \geq 4$.  Notice that for $H_0/m$ large,
the amplitude increases approximately as $(H_0/m)^3$, which will place 
strong restrictions on $\lambda$ in such models.

We remark that we have not included the small corrections to the dynamics
of the effective zero mode and the scale factor arising from the
non-linearities. We have found numerically that these nonlinearities 
are only significant for the
modes that cross about 60 e-folds before the end of inflation for
values of the Hubble parameter $H_0/m_R > 5$.  The effect of these
non-linearities in the large $N$ limit is to slow somewhat the exponential
growth of these modes, with the result of shifting the power spectrum
closer to an exact Harrison-Zeldovich spectrum with $n_s =1$.  Since
for $H_0/m_R > 5$ the power spectrum given by (\ref{index}) differs from
one by at most a few percent, the effects of the non-linearities are
expected to be observationally unimportant.
The spectrum given by (\ref{amplitude}) is
similar to that obtained in references\cite{turner,guthpi} although
the amplitude differs from that obtained there. In addition, we do not
assume slow roll for which $(\nu - \frac{3}{2})\ll 1$, although 
this would be the case if $N_e \gg 60$.

We emphasize an important
feature of the spectrum: it has more power at {\em long
wavelengths} because $\nu-3/2 > 0$. This is recognized to be a
consequence 
of the spinodal instabilities that result in the growth of long wavelength
modes and therefore in more power for these modes.  
This seems to be a robust prediction of new inflationary scenarios in
which the potential has negative second derivative in the region of field
space that produces inflation.  

 It is at this
stage that we recognize the consistency of our approach for separating
the composite effective zero mode from the small fluctuations. We have
argued above that many more than 60 e-folds are required for consistency,
and that the small fluctuations correspond to those modes that cross
the horizon during the last 60 e-folds of the inflationary stage. For these
modes $H_0/k = e^{-H_0 t^*(k)}$ where $t^*(k)$ is the time since the beginning 
of inflation of horizon crossing of the mode with wavevector $k$. 
The scale that  corresponds to the Hubble radius today $\lambda_0
=2\pi/k_0$ is the first to cross during the last 60 or so e-folds
before the end of 
inflation. Smaller scales today will correspond to $k > k_0$ at the
onset of inflation since they will cross the first horizon later and
therefore will reenter earlier. The bound on $|\delta_{k_0}| \propto  
\Delta T/ T \leq  10^{-5}$ on
these scales provides a lower bound on the number of e-folds required for
these type of models to be consistent:
\begin{equation}
N_e >
60+\frac{12}{\nu-\frac{3}{2}}-\frac{\ln(\nu-\frac{3}{2})}{\nu-\frac{3}{2}}\; ,
\label{numbofefolds}
\end{equation}
where we have written the total number of e-folds as $N_e=H_0\; t^*(k_0)+60$.
This in turn can be translated into a bound on the coupling constant using
the estimate given by eq.(\ref{efolds}).

The four year COBE  DMR Sky Map\cite{gorski} gives $n \approx 1.2 \pm 0.3$
thus providing an upper bound on $\nu$
\begin{equation}
0 \leq \nu-\frac{3}{2} \leq 0.05 \label{cobebound}
\end{equation}
corresponding to $h_0 \geq 2.6$. We then find that these values of $h_0$ and
$\lambda \approx 10^{-12}-10^{-14}$ provide sufficient e-folds to satisfy
the constraint for scalar density perturbations. 

\subsection{Tensor Perturbations:} 
The scalar field does not couple to the tensor (gravitational wave)
modes directly, and the tensor perturbations are gauge invariant from
the beginning. Their dynamical evolution is completely determined by
the dynamics of the scale factor\cite{mukhanov,grishchuk2}. 
Having established numerically that the inflationary epoch is
characterized by $\dot{H}/H^2_0 \ll 1$ and that scales of cosmological
interest cross the 
horizon during the stage in which this approximation is excellent, we can
just borrow the known result for the power spectrum of gravitational waves
produced during inflation extrapolated to the matter
era\cite{mukhanov,grishchuk2} 
\begin{equation}
{\cal P}_T(k) \approx \frac{H^2_0}{M^2_{Pl}}k^0\; .
\end{equation}
Thus the spectrum to this order is scale invariant (Harrison-Zeldovich)
with an amplitude of the order $m^4/\lambda M^4_{Pl}$. Then, for values
of $m \approx 10^{12}-10^{14} \mbox{ Gev }$ and 
$\lambda \approx 10^{-12}-10^{-14}$
one finds that the amplitude is $\leq 10^{-10}$ which is much smaller than the
amplitude of scalar density perturbations. 
As usual the amplification of scalar perturbations
is a consequence of the equation of state during the inflationary epoch.

\section{Contact with the Reconstruction Program:}
The program of reconstruction of the inflationary potential seeks to
establish a relationship between features of the inflationary scalar
potential and the spectrum of scalar and tensor perturbations.
This program, in combination with measurements of scalar and tensor
components either from refined measurements of temperature inhomogeneities
of the CMB or through galaxy correlation functions will then offer a 
glimpse of the possible realization of the 
inflation\cite{reconstruction,lyth}. 
Such a reconstruction program is based on the slow roll approximation
and the spectral index of scalar and tensor perturbations are obtained
in a perturbative expansion in the slow roll
parameters\cite{reconstruction,lyth} 
\begin{eqnarray}
\epsilon(\phi) & = &
\frac{\frac{3}{2}\dot{\phi}^2}{\frac{\dot{\phi}^2}{2}+V(\phi)}\; ,
\label{epsifi}  \\
\eta(\phi) & = & -\frac{\ddot{\phi}}{H \dot{\phi}}\; . \label{etafi}
\end{eqnarray}
We can make contact with the reconstruction program by identifying $\phi$
above with our $\phi_{eff}$ after the first few e-folds of inflation needed
to assemble the effective zero mode from the quantum
fluctuations. We have numerically established that for the weak scalar
coupling required 
for the consistency of these models, the cosmologically interesting scales
cross the horizon during the epoch in which $H \approx H_0 \; ; \;
\dot{\phi}_{eff} \approx (\nu - 3/2)\; H_0 \; \phi_{eff} \; ; \; V \approx
m_R^4/\lambda \gg \dot{\phi}^2_{eff}$. In this case we find 
\begin{equation}
\eta(\phi_{eff}) = -(\nu - \frac{3}{2}) \; ; \;  \epsilon(\phi_{eff}) \approx
{\cal O}(\lambda) \ll  \eta(\phi_{eff}).
\end{equation}

With these identifications, and in the notation of\cite{reconstruction,lyth}
the reconstruction program predicts  the index for scalar density
perturbations $n_s$ given by
\begin{equation}
 n_s-1 = -2\left(\nu - \frac{3}{2}\right)+ {\cal O}(\lambda),
\end{equation}
which coincides with the index for  the power
spectrum per logarithmic interval $|\delta_k|^2$ with $|\delta_k|$
given by eq.(\ref{amplitude}).  
We must note however that our treatment did not
assume slow roll for which $(\nu - \frac{3}{2})\ll 1$. Our
self-consistent, non-perturbative study of the dynamics plus the
underlying 
requirements for the identification of a composite operator acting as an
effective zero mode, validates the reconstruction program in weakly
coupled new inflationary models.

\section{DECOHERENCE: QUANTUM TO CLASSICAL TRANSITION DURING INFLATION} 

An important aspect of cosmological perturbations is that 
they are of quantum origin but eventually they become classical as
they are responsible for the small classical metric perturbations. This
quantum to classical crossover is associated with a decoherence process and
has received much attention\cite{polarski,salman}. 

Recent work on decoherence focused on the description of the evolution of the
density matrix for a free scalar massless field that represents the ``velocity
field''\cite{mukhanov} associated with scalar density perturbations\cite{polarski}.  In this section we study
the quantum to classical transition of superhorizon modes for the Bardeen 
variable by relating these to the field mode functions and analyzing the full
time evolution of the density matrix of the matter field. This is accomplished
with the identification given by  equation (\ref{bard}) which relates the mode
functions of the Bardeen variable with those of the scalar field. This relation
establishes that  in the models under consideration the classicality of the Bardeen variable is determined by the
classicality of the scalar field modes.

In the situation under consideration, long-wavelength field modes become
spinodally unstable and grow exponentially after horizon crossing. The 
factorization (\ref{factor}) results in the phases of these modes ``freezing
out''. This feature and the growth in amplitude entail that these modes become
classical. The relation (\ref{bard}) in turn implies that these features 
also apply to the superhorizon modes of the  Bardeen potential. 

 Therefore we can address the quantum
to classical transition of the Bardeen variable (gravitational potential) by
analyzing the evolution of the density matrix for the matter field. 

To make contact with previous work\cite{polarski,salman} we choose to study
the evolution of the field density matrix in conformal time, although the
same features will be displayed in comoving time.

The metric in conformal time takes the form
\begin{equation}\label{metrica}
ds^2= C^2({\cal T})(d{\cal T}^2 - d\vec{x}^2). 
\end{equation}

Upon a conformal rescaling of the field
\begin{equation}\label{camco}
\vec{\Phi}(\vec x, t) = \vec{\chi}(\vec x, {\cal T})/ C({\cal T})\label{confres},
\end{equation}
the action for a scalar field  becomes, after an integration by parts and
 dropping a surface term
\begin{equation}\label{acomf}
S= \int d^3x \, d{\cal T} \left\{\frac12 (\vec{\chi}')^2-\frac12 (\vec{\nabla}\vec{\chi})^2-
{\cal{V}}(\vec{\chi})\right\},
\end{equation}
with
\begin{equation}
{\cal{V}}(\vec{\chi}) =
C^4({\cal T})\, V\left({{\vec{\chi}}\over{C({\cal T})}}\right)-C^2({\cal T})\,
\frac{{\cal{R}}}{12}\, \vec{\chi}^2, 
\end{equation}
where ${\cal{R}}= 6C''({\cal T})/C^3({\cal T})$ is the Ricci scalar,
and primes stand for derivatives with respect to conformal time ${\cal T}$
(for more details see the appendix of ref.\cite{frw2}).
As we can see from eq.(\ref{acomf}), the action takes the same form as
in Minkowski space-time with a modified potential $ {\cal{V}}(\vec{\chi}) $.

The conformal time Hamiltonian operator, which is the generator of translations
in ${\cal T}$, is given by
\begin{equation}
H_{{\cal T}}= \int d^3x \left\{ \frac{1}{2}\Pi^2_{\vec{\chi}}+\frac{1}{2}
(\vec{\nabla}\vec{\chi})^2+{\cal{V}}(\vec{\chi}) \right \}, \label{confham}
\end{equation}
with $\vec{\Pi}_{\chi}$ being the canonical momentum conjugate to $\vec{\chi}$,
$\vec{\Pi}_{\chi} = \vec{\chi}'$. 
Separating the zero mode of the field $\vec{\chi}$ 
\begin{equation}
\vec{\chi}(\vec x, {\cal T}) = \chi_0({\cal T})\delta_{i,1} + {\hat{\vec{\chi}}}(\vec x,{\cal T}),
\end{equation}
and performing the large $N$ factorization on the
fluctuations we find 
that the Hamiltonian becomes linear plus quadratic in the fluctuations, and
similar to a Minkowski space-time Hamiltonian with a ${\cal T}$ dependent mass term given by
\begin{equation}
{\cal{M}}^2({\cal T}) = C^2({\cal T}) \left[m^2+\left(\xi-\frac{1}{6}\right)\,{\cal{R}}
+ \frac{\lambda}{2}\,\chi_0^2({\cal T}) + \frac{\lambda}{2}\,
\langle \hat{\chi}^2 \rangle\right]. \label{masseff}
\end{equation}

We can now follow the steps and use the results of ref.\cite{frwpaper} for the
conformal time evolution of the density matrix by setting $a(t)=1$ in the
proper equations of that reference and replacing the frequencies by
\begin{equation}
\omega^2_k({\cal T}) = \vec{k}^2 + {\cal{M}}^2({\cal T})\; . \label{freqs}
\end{equation}
The expectation value in Eq.(\ref{masseff}) and that of the energy momentum 
tensor are obtained in this
${\cal T}$ evolved density matrix.
[As is clear, we obtain in this way the self-consistent dynamics in the curved cosmological background (\ref{metrica})].

The time evolution of the kernels in the density matrix (see \cite{frwpaper})
is determined by the mode functions that obey
\begin{equation}
\left[ \frac{d^2}{d{\cal T}^2}+k^2+{\cal{M}}^2({\cal T})\right] F_k({\cal T})=0.
\label{fmodeqn}
\end{equation}
The Wronskian of these mode functions
\begin{equation}\label{wff}
{\cal{W}}(F,F^*)= F'_k F^*_k-F_k F'^{*}_k
\end{equation} 
is a constant. It is natural to impose initial conditions such that at the
initial ${\cal T}$ the density matrix  describes a pure state which is the
instantaneous ground state of the  Hamiltonian at this initial time. 
 This implies that the initial conditions of the mode functions
$F_k({\cal T})$ be chosen to be (see \cite{frwpaper})
\begin{equation}
F_k({\cal T}_o)= \frac{1}{\sqrt{\omega_k({\cal T}_o)}} \; \; ; \; 
F'_k({\cal T}_o)= -i\omega_k({\cal T}_o) \;  F_k({\cal T}_o). \label{inicond}
\end{equation}
With such initial conditions, the Wronskian (\ref{wff}) takes the value
\begin{equation}\label{wro}
{\cal{W}}(F,F^*)= -2i \; .
\end{equation}

The Heisenberg field operators $\hat{\chi}(\vec x, {\cal T})$  and 
their canonical momenta $\Pi_{\chi}(\vec x, {\cal T})$ can now be expanded as:
\begin{eqnarray}
&& \hat{\vec{\chi}}(\vec x, {\cal T}) = \int {{d^3k}\over {(2\pi)^{3/2}}} 
\left[ \vec{a}_{ \vec k} \; F_k({\cal T})+ \vec{a}^{\dagger}_{ -\vec k} \;F^*_k({\cal T}) \right]
 e^{i \vec k \cdot \vec x}, \label{heisop}\\ 
&& \vec{\Pi}_{\chi}(\vec x, {\cal T}) =  
\int {{d^3k}\over {(2\pi)^{3/2}}} 
\left[ \vec{a}_{ \vec k}  \;F'_k({\cal T})+ \vec{a}^{\dagger}_{ -\vec k}
\;F'^{*}_k({\cal T}) \right] 
 e^{i \vec k \cdot \vec x}, \label{canheisop}
\end{eqnarray}
with the time independent creation and
annihilation operators $ \vec{a}_{ \vec k} $ and $ \vec{a}^{\dagger}_{ \vec k} $ 
obeying canonical commutation relations. Since the fluctuation fields
in comoving and conformal time are related by a conformal rescaling
given by eq.~(\ref{confres})
it is straightforward to see that the mode functions in comoving time
$ t $ are related to those in conformal time simply as
\begin{equation}
f_k(t) = \frac{F_k({\cal T})}{C({\cal T})}.
\end{equation}
Therefore the initial conditions given in Eq. (\ref{inicond}) on the conformal time mode
functions and the choice $a(t_0)= C({\cal T}_0)=1$  imply the initial conditions for the mode functions in comoving time given by Eq. (\ref{initcond}).

In the large $N$ or Hartree (also in the self-consistent one-loop)
approximation, the density matrix 
is Gaussian, and defined by a normalization factor, a complex covariance that
determines the diagonal matrix elements, and a real covariance that determines
the mixing in the Schr\"odinger representation as discussed in
ref.\cite{frwpaper} (and references therein).

That is, the density matrix takes the form
\begin{eqnarray}\label{matden}
\rho[\Phi,\tilde{\Phi},{\cal T}] & = & \prod_{\vec{k}} {\cal{N}}_k({\cal T})
\exp\left\{ 
- \frac12 A_k({\cal T}) \; \vec{\eta}_{\vec{k}}({\cal T})\cdot
\vec{\eta}_{-{\vec{k}}}({\cal T})- 
\frac12 A^*_k({\cal T}) \;
\tilde{\vec{\eta}}_{\vec{k}}({\cal T})\cdot 
\tilde{\vec{\eta}}_{-{\vec{k}}}({\cal T}) 
\right. \nonumber \\  &   & \left.
- \;  B_k({\cal T}) \; \vec{\eta}_{\vec{k}}({\cal
T})\cdot\tilde{\vec{\eta}}_{-{\vec{k}}}({\cal T}) 
+i\, \vec{\pi}_{\vec{k}}({\cal T})\cdot\left(\vec{\eta}_{-{\vec{k}}}({\cal T})-
\tilde{\vec{\eta}}_{-{\vec{k}}}({\cal T})\right) \right\} ;\ , \\
\vec{\eta}_{\vec{k}}({\cal T})          & = &
\vec{\chi}_{\vec{k}}({\cal T})-\chi_0({\cal T})\,\delta_{i,1} \;\delta(\vec{k}) 
\; \; ; \;  \; 
\tilde{\vec{\eta}}_k({\cal T})        = 
{\tilde{\vec{\chi}}}_k({\cal T})-\chi_0({\cal T})\, 
\delta_{i,1}\; \delta(\vec{k}) \; . \nonumber
\end{eqnarray}
$\vec{\pi}_{\vec{k}}({\cal T}) $ is the Fourier 
transform of $ \Pi_{\chi}({\cal T},\vec{x}) $. This form of the density matrix
is dictated by the hermiticity condition 
$$
\rho[\Phi,\tilde{\Phi},{\cal T}]
=\rho^*[\tilde{\Phi},\Phi,{\cal T}]\; ;
$$
as a result of this, $ B_k({\cal T}) $ is real.
The kernel $ B_k({\cal T}) $ determines the amount of `mixing' in the
density matrix since if $ B_k=0 $, the density matrix corresponds to a pure
state because it factorizes into  a wave functional depending only on
$ \Phi(\cdot) $ times its complex conjugate taken at $ \tilde{\Phi}(\cdot) $. This is
the case under consideration, since the initial conditions correspond to a
pure state and under time evolution the density matrix remains that of a pure
state\cite{frwpaper}. 

In conformal time quantization and in the Schr\"odinger representation in which
the field $\chi$ is diagonal the conformal time evolution of the density matrix
is via the conformal time Hamiltonian (\ref{confham}). The evolution equations
for the covariances is obtained from those given in ref.\cite{frwpaper} by
setting $a(t) = 1$ and using the frequencies $\omega^2_k({\cal T})=
k^2+{\cal{M}}^2({\cal T})$. In particular, by setting the covariance of the
diagonal elements (given by equation (2.20) in \cite{frwpaper}; see
also equation (2.44) of \cite{frwpaper}),
\begin{equation}\label{FpF}
A_k({\cal T}) = -i \, \frac{F'^*_k({\cal T})}{F^*_k({\cal T})}.
\end{equation}

More explicitly \cite{frwpaper},
\begin{eqnarray}\label{coero}
{\cal{N}}_k({\cal T}) &=&{\cal{N}}_k({\cal T}_0)\;
\exp\left[\int_{{\cal T}_0}^{{\cal T}} A_{Ik}({\cal T}')\;d{\cal T}' \right] = 
{{{\cal N }_k({\cal T}_0)} \over { \sqrt{\omega_k({\cal T}_o)} \; |F_k({\cal T})|}} 
\; , \cr \cr
A_{Ik}({\cal T}) &=& - {d \over {d{\cal T}}}\log|F_k({\cal T})| = -{\dot a}(t)-
a(t)\;{d \over {dt}}\log|f_k(t)| \; , \cr \cr
A_{Rk}({\cal T}) &=& {1 \over { |F_k({\cal T})|^2}} \, = { 1\over 
{ a(t)^2 \; |f_k(t)|^2 }} \; ,  \\ \cr
B_k({\cal T}) &\equiv & 0 \; , \nonumber
\end{eqnarray}
where $A_{Rk}$ and $A_{Ik}$ are respectively the real and imaginary parts
of $A_k$ and we have used the value of the Wronskian (\ref{wro}) in
evaluating (\ref{coero}).

The coefficients $ A_k({\cal T}) $ and $ {\cal{N}}_k({\cal T}) $
in the gaussian density matrix (\ref{matden}) are
completely determined by  the conformal mode functions $ F_k({\cal T}) $
(or alternatively the comoving time mode functions $f_k(t)$).

Let us study the time behavior  of these coefficients. 
During inflation, $ a(t) \approx  e^{h_0t}$,
and the mode functions  factorize after horizon crossing, and superhorizon
modes  grow in cosmic time  as in 
Eq.~(\ref{factor}):
$$
a^2(t) |f_k(t)|^2 \approx { 1 \over  {\cal D}_k} \; e^{(2\nu -1)h_0t}
$$ 
where the coefficient $ {\cal D}_k $ can be read from eq.~(\ref{factor}).

We emphasize that this is a {\em result of the full evolution} as displayed from the numerical solution in fig. (\ref{modu}). These mode functions
encode all of the self-consistent and non-perturbative features of the
dynamics. This should be contrasted with other studies in which typically
free field modes in a background metric are used.  

 Inserting this expression in eqs.(\ref{coero}) yields, 
\begin{eqnarray}\label{roasi}
A_{Ik}({\cal T})&\buildrel{t \to \infty}\over =&-h_0 \;  e^{h_0t} \left(\nu -
\frac12 \right)+ {\cal O}(e^{-h_0t})    \; , \cr \cr
A_{Rk}({\cal T}) &\buildrel{t \to \infty}\over =& {\cal D}_k\; \; 
 e^{-(2\nu -1)h_0t}  . \nonumber
\end{eqnarray}

Since $ \nu -\frac12 > 1 $,
we see that the imaginary part of the covariance $  A_{Ik}({\cal T}) $
{\it grows} very fast. Hence, the off-diagonal elements of $
\rho[\Phi,\tilde{\Phi},{\cal T}] $ oscillate wildly after a few e-folds of
inflation. In particular their contribution to expectation values of operators
will be washed out. That is, we quickly reach a {\it classical} regime where
only 
the diagonal part of the density matrix is relevant:
\begin{equation}\label{claden}
\rho[\Phi,\Phi,{\cal T}] = \prod_{\vec{k}} {\cal{N}}_k({\cal T})
\exp\left\{ -  A_{Rk}({\cal T}) 
\; \eta_{\vec{k}}({\cal T})\;{\eta}_{-{\vec{k}}}({\cal T}) \right\}.
\end{equation}

The real part of the covariance $ A_{Rk}({\cal T}) $ (as well as any non-zero
mixing kernel $ B_k({\cal T}) $\cite{frwpaper}) {\it decreases} as $ e^{-(2\nu
-1)h_0t} $. Therefore, characteristic field configurations $ \eta_{\vec k} $
are very large (of order $ e^{(\nu -\frac12)h_0t} $). Therefore configurations
with field amplitudes up to ${\cal O}(e^{(\nu -\frac12)h_0t})$ will have a
substantial probability of occurring and being represented in the density
matrix.

Notice that $ \chi \sim  e^{(\nu -\frac12)h_0t} $  corresponds to
field configurations $ \Phi $ with amplitudes of order  $  e^{(\nu -\frac32)h_0t} $ [see eq.~(\ref{camco})]. It is the fact that $ \nu -\frac32 > 0 $ which in this
situation is responsible for the ``classicalization'', which is  seen to
be a consequence of the spinodal growth of long-wavelength fluctuations.

The equal-time field correlator is given by 
\begin{eqnarray}
\langle \bar{\chi}(\vec x,{\cal T})\; \bar{\chi}({\vec x}',{\cal T}) \rangle 
&=& \int \frac{d^3k}{2(2\pi)^3}\; |F_k({\cal T})|^2\; e^{i{\vec k}.({\vec x}-
{\vec x}')} \quad   , \cr \cr
& = & a(t)^2 \; \int \frac{d^3k}{2(2\pi)^3}\; 
|f_k(t)|^2\; e^{i{\vec k}.({\vec x}-{\vec x}')} \quad .
\end{eqnarray}
and is seen to be dominated by the superhorizon mode functions and to grow as 
$  e^{(2\nu -1)h_0 t} $, whereas the
field commutators remain fixed showing the emergence of a classical behavior.
As a result we obtain
\begin{equation}
\langle \bar{\chi}(\vec x,{\cal T})\; \bar{\chi}({\vec x}',{\cal T}) \rangle 
\propto a^2(t) \; \phi_{eff}(t) \;\phi_{eff}(t') \; G(|\vec x - \vec{x}'|)+
\mbox{ small }
\end{equation}
where $G(|\vec x - \vec{x}'|)$ falls off exponentially for distances larger than
the horizon\cite{De Sitter} and ``small'' refers to terms that are 
smaller in magnitude. This factorization of the correlation functions
is another indication of classicality.

Therefore, it is possible to  describe the physics by using
classical field theory. 
More precisely, one can use a classical statistical (or stochastic) field theory
described by the functional probability distribution (\ref{claden}).

These results generalize 
the decoherence treatment given in  ref.\cite{ps}
for a free massless field in pure quantum states to the case of interacting
fields with broken symmetry.
Note that the formal decoherence or classicalization in the density matrix
appears after the modes with wave vector $k$ become superhorizon sized i.e. 
when the factorization of the mode functions becomes effective.  

\section{Conclusions:} 

It can be argued that the inflationary paradigm as currently understood is one
of the greatest applications of quantum field theory. The imprint of quantum
mechanics is everywhere, from the dynamics of the inflaton, to the generation of
metric perturbations, through to the reheating of the universe. It is clear
then that we need to understand the quantum mechanics of inflation in as deep a
manner as possible so as to be able to understand what we are actually testing
via the CMBR temperature anisotropies, say.

What we have found in our work is that the quantum mechanics of inflation is
extremely subtle. We now understand that it involves both non-equilibrium as
well as non-perturbative dynamics and that what you start from may {\it not} be
what you wind up with at the end!

In particular, we see now that the correct interpretation of the
non-perturbative growth of quantum fluctuations via spinodal decomposition is
that the background zero mode must be redefined through the process of zero
mode reassembly that we have discovered. When this is done (and {\it only}
when!) we can interpret inflation in terms of the usual slow-roll approach with
the now small quantum fluctuations around the redefined zero mode driving the
generation of metric perturbations. 

We have studied the non-equilibrium dynamics of a `new inflation' scenario in a
self-consistent, non-perturbative framework based on a large $ N $ expansion,
including the dynamics of the scale factor and backreaction of quantum
fluctuations. Quantum fluctuations associated with superhorizon modes grow
exponentially as a result of the spinodal instabilities and contribute to the
energy momentum tensor in such a way as to end inflation consistently.

Analytical and numerical estimates have been provided that establish the regime
of validity of the classical approach.  We find that these superhorizon modes
re-assemble into an effective zero mode and unambiguously identify the
composite field that can be used as an effective expectation value of the
inflaton field whose {\em classical} dynamics drives the evolution of the scale
factor. This identification also provides the initial condition for this
effective zero mode.

A consistent criterion is provided to extract ``small'' fluctuations that will
contribute to cosmological perturbations from ``large'' non-perturbative
spinodal fluctuations. This is an important ingredient for a consistent
calculation and interpretation of cosmological perturbations.  This criterion
requires that the model must provide many more than 60 e-folds to identify the
`small perturbations' that give rise to scalar metric (curvature)
perturbations. We then use this criterion combined with the gauge invariant
approach to obtain the dynamics of the Bardeen variable and the spectrum for
scalar perturbations.

We find that during the inflationary epoch, superhorizon modes of the Bardeen
potential grow exponentially in time reflecting the spinodal
instabilities. These long-wavelength instabilities are manifest in the spectrum
of scalar density perturbations and result in an index that is less than
one, i.e. a ``red'' power spectrum, providing more power at long wavelength.  We
argue that this `red' spectrum is a robust feature of potentials that lead to
spinodal instabilities in the region in field space associated with inflation
and can be interpreted as an ``imprint'' of the phase transition on the
cosmological background. Tensor perturbations on the other hand, are not
modified by these features, they have much smaller amplitude and a
Harrison-Zeldovich spectrum.

We made contact with the reconstruction program and validated the results for
these type of models based on the slow-roll assumption, despite the fact that
our study does not involve such an approximation and is non-perturbative.

Finally we have studied the quantum to classical crossover and decoherence of
quantum fluctuations by studying the full evolution of the density matrix, thus
making contact with the concept of ``decoherence without
decoherence''\cite{polarski} which is generalized to the interacting case. In
the case under consideration decoherence and classicalization is a consequence
of spinodal growth of superhorizon modes and the presence of a horizon. The
phases of the mode functions ``freeze out'' and the amplitudes of the
superhorizon modes grow exponentially during the inflationary stage, again as a
result of long-wavelength instabilities.  As a result field configurations with
large amplitudes have non-vanishing probabilities to be represented in the
dynamical density matrix.  In the situation considered, the quantum to
classical crossover of cosmological perturbations is directly related to the
``classicalization'' of superhorizon matter field modes that grow exponentially
upon horizon crossing during inflation. The diagonal elements of the density
matrix in the Schroedinger representation can be interpreted as a classical
distribution function, whereas the off-diagonal elements are strongly
suppressed during inflation.

\section{Acknowledgements:} 
The authors thank J. Baacke, K. Heitman, L. Grishchuk, E. Weinberg, E. Kolb,
A. Dolgov and D. Polarski, for conversations and discussions. D. B. thanks the
N.S.F for partial support through the grant awards: PHY-9605186 and
INT-9216755, the Pittsburgh Supercomputer Center for grant award No: PHY950011P
and LPTHE for warm hospitality.  R. H., D. C. and S. P. K. were supported by
DOE grant DE-FG02-91-ER40682.  The authors  acknowledge partial support by
NATO.

\end{document}